\documentclass[5p]{elsarticle}

\usepackage{lineno}
\usepackage{hyperref}
\modulolinenumbers[5]

\journal{Journal of \LaTeX\ Templates}

\usepackage{amsmath}
\usepackage{fleqn}
\usepackage[caption=false,font=normalsize,labelfont=sf,textfont=sf]{subfig}

\usepackage[table]{xcolor}
\usepackage{multirow}
\usepackage{hhline}
\usepackage{rotating}
\usepackage{fixltx2e}

\usepackage{cleveref}

\begin{document}

\begin{frontmatter}

\title{Adversarial dictionary learning for a robust analysis and modelling of spontaneous neuronal activity}
\tnotetext[mytitlenote]{Fully documented templates are available in the elsarticle package on \href{http://www.ctan.org/tex-archive/macros/latex/contrib/elsarticle}{CTAN}.}


\author[addr1,addr2]{Eirini Troullinou}
\author[addr2]{Grigorios Tsagkatakis}
\author[addr3,addr4]{Ganna Palagina}
\author[addr1,addr2]{Maria Papadopouli}
\author[addr3,addr4]{Stelios Manolis Smirnakis}
\author[addr1,addr2]{Panagiotis Tsakalides}

\address[addr1]{Department of Computer Science, University of Crete, Heraklion, 70013, Greece}
\address[addr2]{Institute of Computer Science, Foundation for Research and Technology Hellas, Heraklion, 70013, Greece}
\address[addr3]{Department of Neurology, Brigham and Women's Hospital, Harvard Medical School,Boston MA 02115}
\address[addr4]{Boston VA Research Institute, Jamaica Plain Veterans Administration Hospital, \\ Harvard Medical School, Boston, United States}

\begin{abstract}
The field of neuroscience is experiencing rapid growth in the complexity and quantity of the recorded neural activity allowing us unprecedented access to its dynamics in different brain areas. The objective of this work is to discover directly from the experimental data rich and comprehensible models for brain function that will be concurrently robust to noise. Considering this task from the perspective of dimensionality reduction, we develop an innovative, robust to noise dictionary learning framework based on adversarial training methods for the identification of patterns of synchronous firing activity as well as within a time lag. We employ real-world binary datasets describing the spontaneous neuronal activity of laboratory mice over time, and we aim to their efficient low-dimensional representation. The results on the classification accuracy for the discrimination between the clean and the adversarial-noisy activation patterns obtained by an SVM classifier highlight the efficacy of the proposed scheme compared to other methods, and the visualization of the dictionary's distribution demonstrates the multifarious information that we obtain from it.
\end{abstract}

\begin{keyword}
Dictionary Learning, Supervised Machine Learning, biological neural networks.
\end{keyword}

\end{frontmatter}


\section{Introduction}

The advances of imaging and monitoring technologies, such as in vivo 2-photon calcium imaging at the mesoscopic regime as well as the massive increases in computational power and algorithmic development have enabled advanced multivariate analyses of neural population activity, recorded either sequentially or simultaneously.

More specifically, high resolution optical imaging methods have recently revealed the dynamic patterns of neural activity across the layers of the primary visual cortex (V1) leading to this important question: Neuronal groups that fire in synchrony may be more efficient at relaying shared information and are more likely to belong to networks of neurons subserving the same function. By using 2-photon imaging, we monitored the spontaneous population bursts of activity in pyramidal cells and interneurons of mouse in L2/3 V1. We found that the sizes of spontaneous population bursts and the degree of connectivity of the neurons in specific fields of view (FOVs) formed scale-free distributions, suggestive of a hierarchical small-world net architecture \cite{palagina2019inhibitory}. The existence of such groups of "linked" units inevitably shapes the profile of spontaneous events observed in V1 networks \cite{miller2014visual,kenet2003spontaneously,luczak2009spontaneous}. Thus, the analysis of the spontaneous activity patterns provides an opportunity for identifying groups of neurons that fire with increased levels of synchrony (have significant "functional connectivity" between each other). 

In order to analyze these populations and to find features that are not apparent at the level of individual neurons, we adopt dictionary learning (DL) methods, which provide a parsimonious description of statistical features of interest via the output dictionary, discarding at the same time some aspects of the data as noise. Moreover, dictionaries are a natural approach for performing exploratory data analysis as well as visualization. Given the fact that the dictionary is the new space of reduced dimensionality, the computational complexity of its management is much smaller compared to the initial raw data and thus, for all these advantages, DL has been applied in various domains. 

In brain signaling specifically, the K-SVD algorithm \cite{aharon2006rm}, has been used for capturing the behavior of neuronal responses into a dictionary, which was evaluated with real-world data for its generalization capacity as well as for its sensitivity with respect to noise  \cite{troullinou2017dictionary}. DL has been also suggested for the EEG (electroencephalography) inverse problem. Specifically, Liu \textit{et al.} \cite{liu2017sparse} proposed a supervised formulation of source reconstruction and supervised source classification to address the estimation of brain sources and to distinguish the various sources associated with different status of the brain. Moreover, accurate EEG signal classification plays an important role in the performance of BCI (Brain Computer Interface) applications. Ameri \textit{et al.} \cite{zhou2018online} adapted the projective dictionary pair learning method (DPL) for EEG signal classification. They learned a synthetic as well as an analysis dictionary, which were used in the classification step to increase the speed and accuracy of the classifier. Morioka \textit{et al.} \cite{morioka2015learning} proposed a dictionary learning, sparse coding method to address the issue of the inherent variability existing in brain signals caused by different physical and mental conditions among multiple subjects and sessions. Such variability complicates the analysis of data from multiple subjects and sessions in a consistent way, and degrades the performance of neural decoding in BCI applications. 

In this work, we propose the Adversarial Dictionary Learning (ADL) algorithm, which captures the synchronicity patterns among neurons, and its extended version, the Relaxed Adversarial Dictionary Learning (RADL) for cofiring patterns within a time lag. Adversarial training is the process of explicitly training a model on adversarial examples, in order to increase its robustness to noisy inputs. Thus, we create an adversarial learning environment by using clean and adversarial-noisy activation patterns. The main objectives are the construction of a dictionary that will be robust to the measurement noise (i.e. calcium fluctuations) as well as to the identification of firing events emerging by chance. Both ADL and RADL construct the output dictionary by selecting only those patterns of the input data that contribute to a better reconstructed representation of the clean input signal than the adversarial-noisy one. After obtaining our trained dictionary, we quantify its quality and robustness by training a supervised classifier with the reconstructed clean and noisy signals as well as with the raw ones, and examine when the classifier exhibits the smallest testing error. To assess whether the trained dictionary has captured the underlying statistics of the input data, we employ the classification accuracy (i.e. the extent to which the classifier can discriminate the clean from the noisy signal).
 
To validate the proposed algorithms, we employed two real-world binary datasets that depict the neuronal activity of a 9-day old and a 36-day old C57BL/6 laboratory mouse. Data was collected using 2-photon calcium imaging in the V1, L2/3 area of the neocortex of the animals. Fig. \ref{fig1} illustrates the format of our data, where each column represents an example-activation pattern that consists of $0s$ for the non-firing events, while $1s$ represent the firing events.  

While DL has delivered impressive results in various domains (such as pattern recognition, and data mining), the construction of the appropriate dictionary depending on the application still remains challenging. A common drawback in DL algorithms is the generation of real-numbered dictionaries, which in our domain have no physical meaning, and thus they cannot be directly used for extracting useful information from the data nor for visualizations. Thus, an innovative aspect of our work is shaped by the requirement of constructing binary dictionaries (given the binary activation patterns). Additionally, while the majority of the algorithms require a dictionary size parameter, often there is no prior-knowledge on the number of patterns that should be used. To overcome these limitations, ADL constructs a dictionary, using the most representative and robust patterns of the input data and automatically estimates the dictionary size, as the algorithm does this itself during the dictionary construction. RADL offers the same benefits and is extended to discover temporal patterns within a lag (i.e. temporal patterns in larger time windows). The contributions of this work are summarized as follows:
\begin{itemize}
\item Adversarial DL outputs robust to noise dictionaries by excluding those patterns from the input data, which could be a result of noise, caused mainly from calcium fluctuations or other sources of imaging noise.
\item Acquisition of an interpretable dictionary, as the dictionary elements are selected from the input data and thus, the dictionary construction is not a result of a mathematical transformation, as opposed to other methods, such as K-SVD \cite{aharon2006rm} or PCA \cite{jolliffe1986principal}.
\item In contrast to other methods that require a choice of dimensionality $K$ (dictionary size), here this is not a parameter that has to be determined by the user, or be estimated (e.g. based on the choice of arbitrary cutoff values or cross-validation methods \cite{cunningham2014dimensionality}).
\item Detection of statistically significant synchronous and within a lag temporal patterns of activity, which can be distinguished from shuffled data (adversarial-noisy examples), whose temporal correlations are destroyed.
\end{itemize}
The remainder of the paper is organized as follows: In Section II, we describe the proposed approaches. Evaluation methodology and experimental results are presented in Section III. Related work is reported in Section IV, while conclusions are drawn in Section V.

\section{Proposed Dictionary Learning Framework}
\begin{figure}[!h] 
\vspace{5pt}
\centerline{\includegraphics[width=0.4\textwidth]{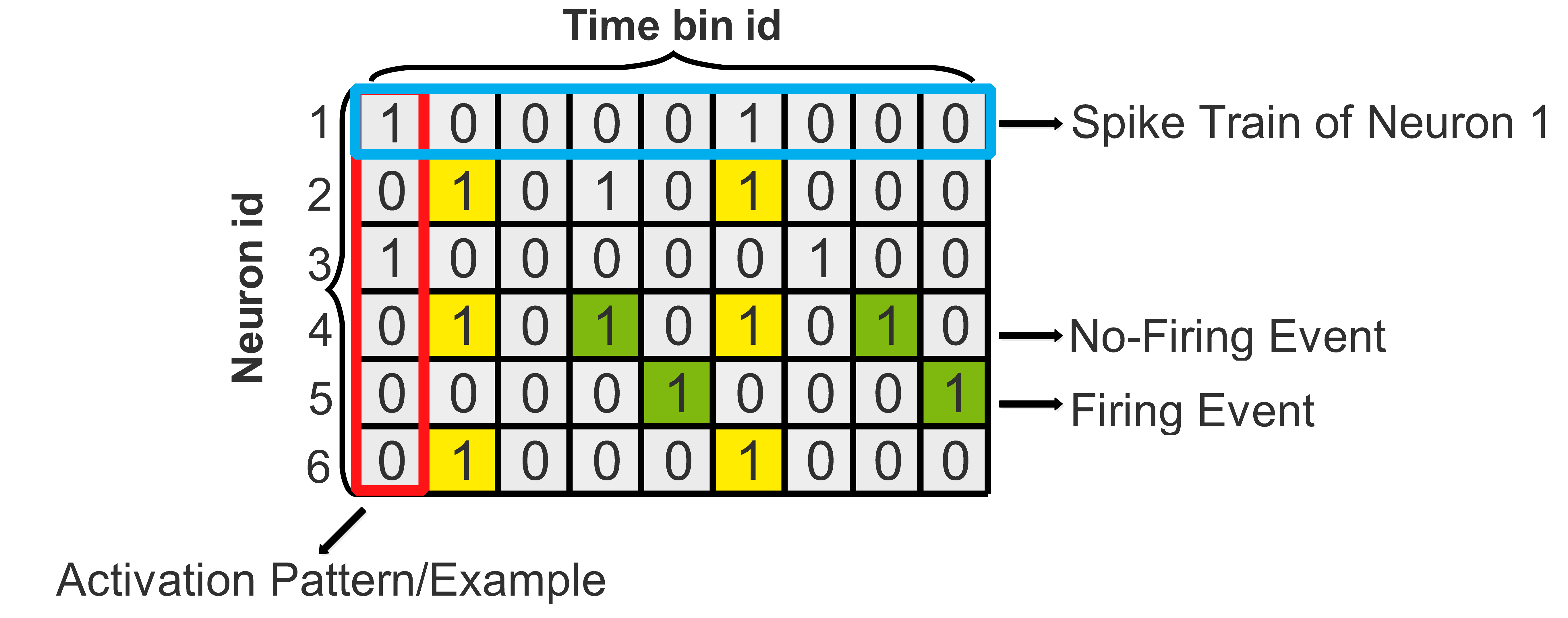}}
\caption{Temporal patterns: Synchronous ($W=1$) and within larger time windows ($W>1$).}
\label{fig1}
\end{figure}

\noindent In this section we present the proposed DL methods: 
\begin{itemize}
\item Adversarial Dictionary Learning Algorithm (ADL) identifies the synchronicity patterns, i.e. patterns where the neurons fire within the same time bin ($W=1$). For example, in Fig. \ref{fig1} Neurons 2, 4 and 6 (yellow boxes) fire simultaneously.
\item Relaxed Adversarial Dictionary Learning Algorithm (RADL) is the extension of ADL, which gives the potential to detect firing activity within a temporal window of length that is determined by the user. For example, in Fig. \ref{fig1} Neurons 4 and 5 (green boxes) are not activated simultaneously but within a time window interval $W=2$. 
\end{itemize}

\noindent We also employ a supervised machine learning framework to quantify the learning capacity of the dictionaries that are produced by the two methods as well as their robustness to adversarial noise.

\subsection{Adversarial Dictionary Learning Algorithm}
ADL aims to identify synchronous activation patterns existing in the input data and outputs them to a dictionary. It is an iterative algorithm, which in every iteration selects randomly an example-activation pattern from the data and examines if it will be included in the dictionary or not. Every iteration consists of two stages. In the first stage, the algorithm examines the contribution of the selected example in the representation of the input data via two representation errors. In the second stage it examines the contribution of the example in the representation of the noisy data (i.e. data that we have artificially added noise) based also on two other representation errors. When these two stages are completed, they are combined in order to determine if the selected example will be included in the dictionary or not. 

Given a training set $\mathbf{Y}_{clean} \in B^{M \times N}$, where $B$ is the binary set consisting of $0$ and $1$, $M$ is the number of neurons, $N$ the number of clean examples $\mathbf({y}_j)_{j=1}^N$, where each one represents an activation pattern (i.e. the activity of all neurons within one time bin as shown in Fig. \ref{fig1}), we aim to construct a dictionary $\mathbf{D} \in B^{M \times K}$, which at the end of the algorithm will have $K$ dictionary elements that capture the activity among those neurons. Zero columns and those with only one 1-entry (firing of only one neuron within one time bin) have been removed from the training set $\mathbf{Y}_{clean}$, as we are interested only in synchronicity patterns (i.e. when two or more neurons fire simultaneously within the same time bin).

ADL constructs the dictionary $\mathbf{D}$ incrementally, as in every iteration of the algorithm one example $\mathbf{y}_i$ of the set $\mathbf{Y}_{clean}$ is examined as to whether it will be included in the dictionary or not. The algorithm iterates $N$ times (i.e. for each one of the examples $\mathbf{y}_j$ that are in the set $\mathbf{Y}_{clean}$ and stops when all of them are examined. Apart from the output dictionary $\mathbf{D}$ the algorithm also uses an auxiliary dictionary $\mathbf{D'}$, which in every iteration of the algorithm has all the elements of $\mathbf{D}$ as well as an extra example $\mathbf{y}_i$, which at the current iteration is the example that is examined whether it will be included in the dictionary $\mathbf{D}$ or not. Namely, if at the iteration $i$, $\mathbf{D} \in B^{M \times k}$ then $\mathbf{D'} \in B^{M \times (k+1)}$. $\mathbf{D}$ is initialized randomly with an example $\mathbf{y}_j$ of the set $\mathbf{Y}_{clean}$ and at the first iteration of the algorithm when the first $\mathbf{y}_i$ is to be examined, dictionaries $\mathbf{D}$ and $\mathbf{D'}$ have the following form:
\begin{align} \label{eq:1}
\mathbf{D=y}_j ~~\text{and}~~ \mathbf{D'}=[\mathbf{D}, \mathbf{y}_i]=[\mathbf{y}_j, \mathbf{y}_i]
\end{align}

At the first stage of the algorithm, in order to validate and decide if the example $\mathbf{y}_i$ should be included in the dictionary or not, we also use a set of clean validation examples $\mathbf{V}_{clean} \in B^{M \times (N-1)}$, which consists of all the examples of set $\mathbf{Y}_{clean}$, except the current example $\mathbf{y}_i$ under consideration, namely $\mathbf{V}_{clean}={\left\lbrace(\mathbf{y}_j)_{j=1}^{N-1}, j\neq{i}\right\rbrace}$. According to the sparse representation framework, given the dictionaries $\mathbf{D}$ and $\mathbf{D'}$, we search respectively for the coefficient matrices $\mathbf{X} \in R^{k \times N}$ and $\mathbf{X'} \in R^{(k+1)\times N}$. An approach to this problem is the minimization of the following $l_0 $ norm problems:
\begin{align} 
\underset{\mathbf{X}} {\operatorname{min}} ||\mathbf{V}_{clean}-\mathbf{D}  \mathbf{X}||_2^2, \ \ \text{subject to} \ \ ||x_j||_0\leq T_0 \label{eq2}\\ 
\underset{\mathbf{X'}} {\operatorname{min}} ||\mathbf{V}_{clean}-\mathbf{D'}\mathbf{X'}||_2^2, \ \ \text{subject to} \ \ ||x'_j||_0\leq T_0 \label{eq3}
\end{align}
where $||{\bf x}_j||_0$ and $||{\bf x'}_j||_0$ are the $l_0$ pseudo-norms, which correspond to the number of non-zero elements for every column $j$ of the sparse coefficient matrices ${\bf X}$ and ${\bf X'}$, respectively.
The sparsity level $T_0$ denotes the maximal number of non-zero elements for every column $j$ of ${\bf X}$ and ${\bf X'}$. Namely each column can have at most $T_0$ elements. These minimization problems are solved using the OMP Algorithm \cite{tropp2007signal}.

Based on \cref{eq2,eq3}, we examine whether ${\bf DX}$ or ${\bf D'X'}$, which represent the sets $\mathbf{V}_{clean\_reconstructed}$ and $\mathbf{V'}_{clean\_reconstructed}$ respectively, better approach the validation set of examples ${\bf V}_{clean}$. Thus, the question under discussion is if the example ${\bf y}_i$, which is included in ${\bf D'}$, contributes to a better representation of the set ${\bf V}_{clean}$. The metric we used to answer this question is:

\begin{align} 
\mathbf{E}_{clean}=\left\lbrace \text{RMSE}(\mathbf{V}_{clean},\mathbf{V}_{clean \textunderscore reconstructed})\right\rbrace \label{eq4}\\
\mathbf{E'}_{clean}=\left\lbrace \text{RMSE}(\mathbf{V}_{clean},\mathbf{V'}_{clean \textunderscore reconstructed})\right\rbrace \label{eq5}
\end{align}
where RMSE is the root mean squared error. In case of
\begin{align}	\label{eq6}
 \mathbf{E'}_{clean}  <  \mathbf{E}_{clean}
\end{align}
this means that the example ${\bf y}_i$, which was only included in ${\bf D'}$ had indeed an effective result in the representation of the validation set $\mathbf{V}_{clean}$.

\begin{figure}[!h] 
\vspace{5pt}
\centerline{\includegraphics[width=0.47\textwidth]{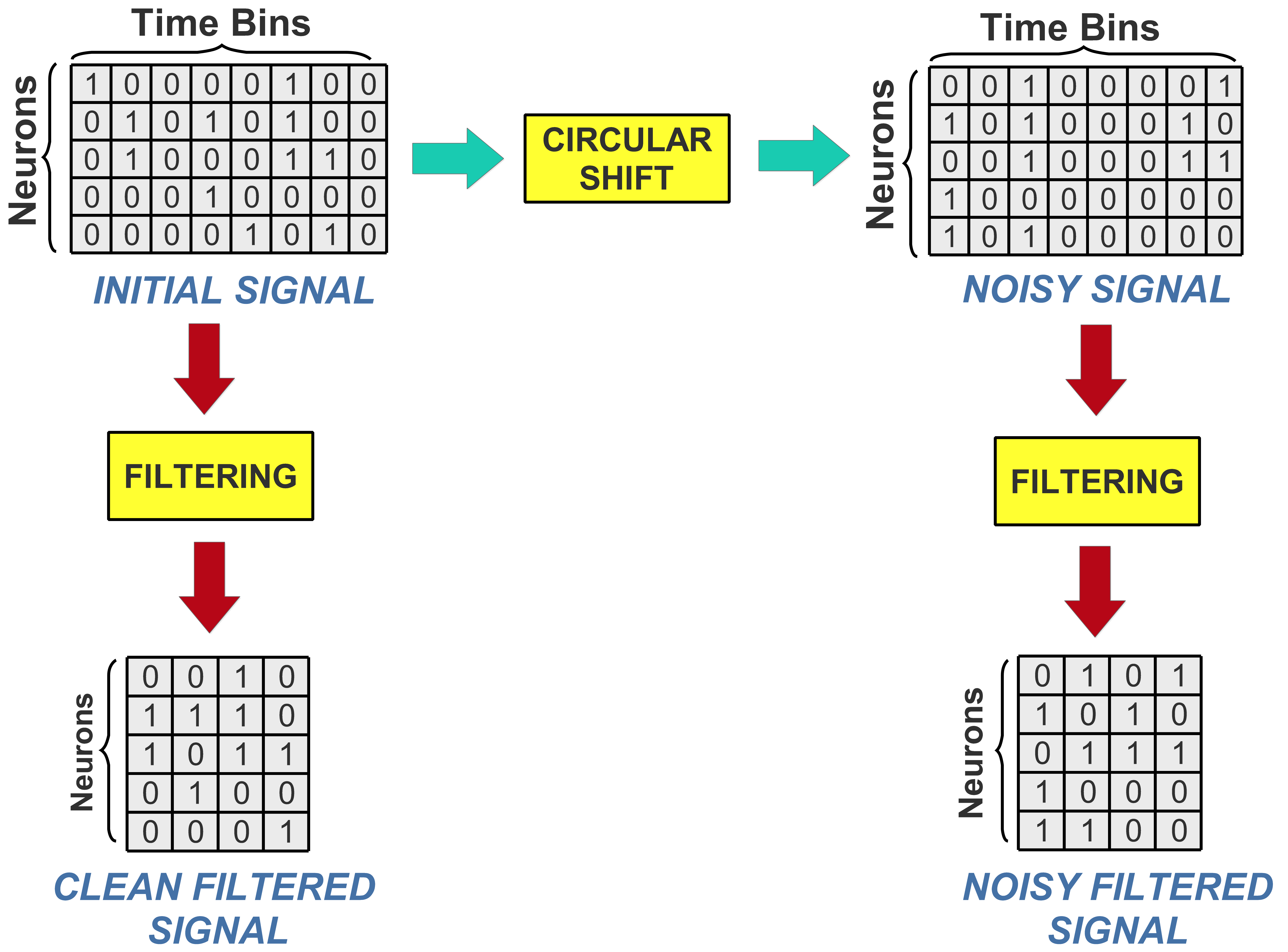}}
\caption{Creation of noisy dataset with circular shift and removal of zero columns and those where only one neuron is active from the initial and the noisy signal (filtering).}
\label{fig2}
\end{figure}

We will keep up with the description of the second stage of our algorithm, which is partially inspired from adversarial learning methods \cite{huang2011adversarial,stone1998towards}, justifying the characterism adversarial that we have given to it. The combination of the first and second stage will determine if the example ${\bf y}_i$ will be ultimately added in dictionary ${\bf D}$. More specifically, in order to include the example ${\bf y}_i$ in dictionary ${\bf D}$, besides its good contribution to the representation of the validation set ${\bf V}_{clean}$, it should be simultaneously a non-helpful factor for the representation of an adversarial noisy signal. This aims to the creation of a dictionary that will be robust to noise. In order to achieve this, we create a set of adversarial-noisy examples $\mathbf{Y}_{noisy} \in B^{M \times N}$ by circularly shuffling the spike train of each neuron of the initial set $\mathbf{Y}_{clean}$ by a random number, different for each neuron. Fig. \ref{fig2} depicts a simple example with five neurons spiking at various time bins showing how the adversarial-noisy signal is created. In order to create the noisy signal, we perform circular shifting to each neuron of the initial signal independently. For example, the spike train of the first neuron is circularly shifted by 2 positions-time units. Accordingly, the spike train of the second neuron is circularly shifted by 5 positions-time units etc. From both the initial and the noisy signal, zero columns and those with one single active neuron are removed (filtering). This type of noise is much more realistic compared to other types, such as random flipping of events, gaussian noise etc., as it preserves the spike distribution of each neuron (firing rate), while it destroys the synchronicity patterns between individual neurons. We also create a validation set of noisy examples ${\bf V}_{noisy} \in B^{M \times (N-1)}$, which consists of all the examples included in set $\mathbf{Y}_{noisy}$ except from a random one that is removed so that ${\bf V}_{clean}$ and ${\bf V}_{noisy}$ have the same number of examples. 
 
In order to evaluate the contribution of the example ${\bf y}_i$ to the representation of the set $\mathbf{V}_{noisy}$, the following minimization problems are solved using again the OMP algorithm: 
\begin{align} 
\underset{\mathbf{X}_{noisy}} {\operatorname{min}} ||\mathbf{V}_{noisy}-\mathbf{D}\mathbf{X}_{noisy}||_2^2, \ \ \text{s.t.} \ \ ||x_{j,noisy}||_0\leq T_0 \label{eq7}\\
\underset{\mathbf{X'}_{noisy}} {\operatorname{min}} ||\mathbf{V}_{noisy}-\mathbf{D'}\mathbf{X'}_{noisy}||_2^2, \ \ \text{s.t.} \ \ ||x'_{j,noisy}||_0\leq T_0 \label{eq8}
\end{align}
Using the same metric as that in \cref{eq4,eq5}, we get the following representation errors:
\begin{align} 
\mathbf{E}_{noisy}=\left\lbrace \text{RMSE}(\mathbf{V}_{noisy},\mathbf{V}_{noisy\textunderscore reconstructed})\right\rbrace \label{eq9}\\
\mathbf{E'}_{noisy}=\left\lbrace \text{RMSE}(\mathbf{V}_{noisy},\mathbf{V'}_{noisy\textunderscore reconstructed})\right\rbrace \label{eq10}
\end{align}
This time we should have
\begin{align} \label{eq11}
\mathbf{E'}_{noisy}  > \mathbf{E}_{noisy}
\end{align}

\begin{figure*}[!ht] 
\centerline{\includegraphics[width=0.9\textwidth]{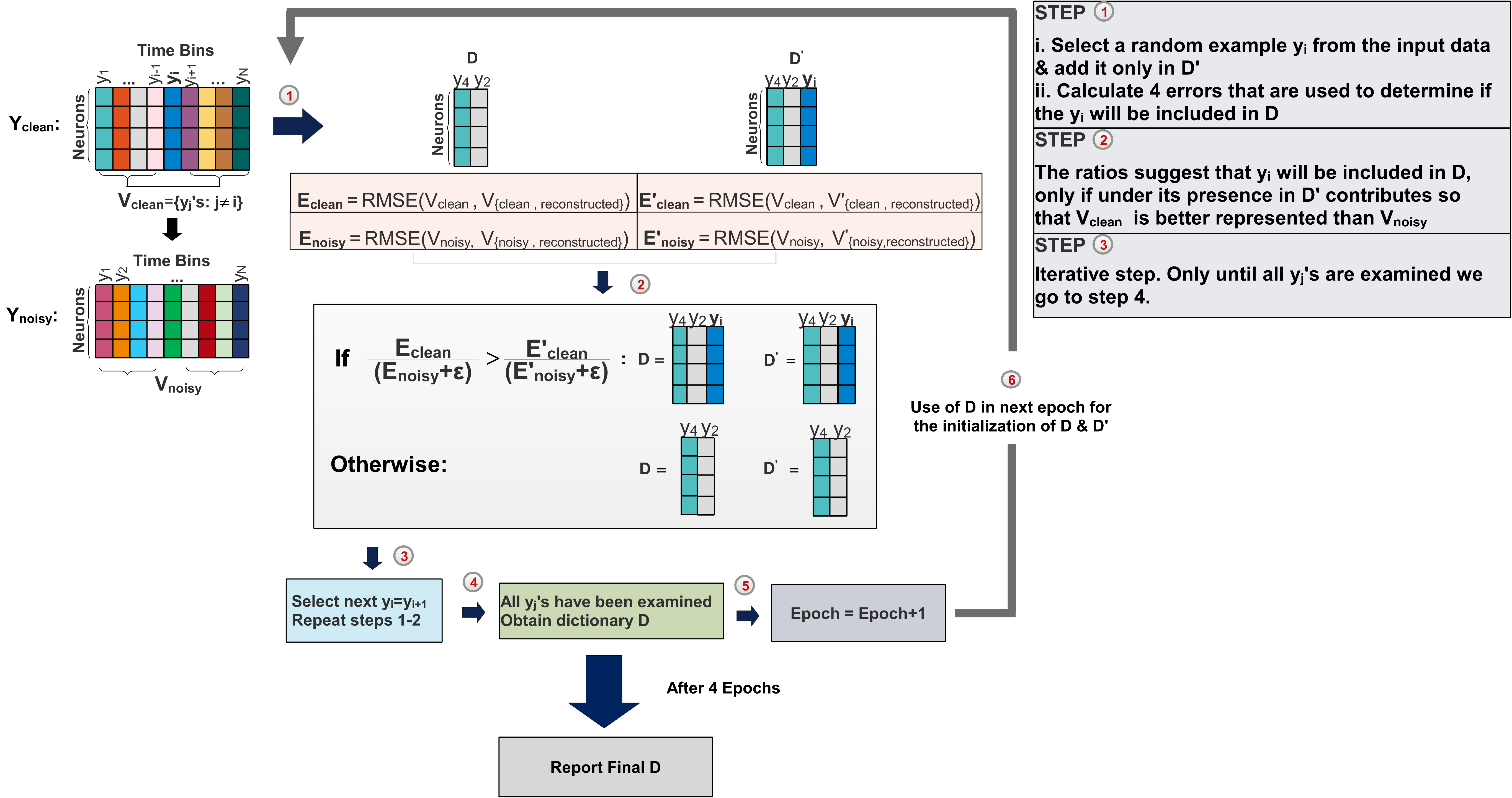}}
\caption{Proposed approach: ADL selects all the appropriate examples of set ${\bf Y}_{clean}$ (steps 1-5) and obtains a dictionary ${\bf D}$. Steps 1-5 are repeated 4 times-epochs and in every epoch dictionaries ${\bf D}$ and ${\bf D'}$ are initialized with the dictionary obtained from the previous epoch (step 6). After the 4 epochs we report the final ${\bf D}$.}
\label{fig3}
\end{figure*}

A bigger error in $\mathbf{E'}_{noisy}$ suggests that the presence of the example ${\bf y}_i$ in dictionary ${\bf D'}$ does not contribute to the good representation of the noisy set of examples ${\mathbf{V}_{noisy}}$. That would be exactly the prerequisite for the inclusion of the example ${\bf y}_i$ in ${\bf D}$, if we took into account only the second part of our algorithm. Note that the dictionary ${\bf D}$ consists only of examples from the set ${\bf Y}_{clean}$. The set ${\bf V}_{noisy}$, which results from the set ${\bf Y}_{noisy}$ is used by the algorithm during the training procedure only in order to determine the appropriateness of the example ${\bf y}_i$ in the dictionary ${\bf D}$.

Eventually, to determine whether or not to include ${\bf y}_i$ in dictionary ${\bf D}$, (\ref{eq6}) and (\ref{eq11}) are combined in the following way: 
\begin{align} \label{eq12}
\frac{\mathbf{E'}_{clean}}{ \mathbf{E'}_{noisy}+\epsilon}<\frac{ \mathbf{E}_{clean}}{ \mathbf{E}_{noisy}+\epsilon}
\end{align}\\
where $\epsilon$ is a very small positive quantity, so as zero denominators are avoided.

If (\ref{eq12}) holds, then ${\bf y}_i$ will be also added in dictionary ${\bf D}$. Dictionaries ${\bf D}$ and ${\bf D'}$ would then temporarily be exactly the same, until the next iteration, where another example ${\bf y}_i$ would be added in dictionary ${\bf D'}$, in order to be examined as to whether it should be eventually included in ${\bf D}$ or not. Otherwise, if 
\begin{align} \label{eq13}
\frac{\mathbf{E'}_{clean}}{ \mathbf{E'}_{noisy}+\epsilon}\geq\frac{ \mathbf{E}_{clean}}{ \mathbf{E}_{noisy}+\epsilon}
\end{align}
then ${\bf y}_i$ is removed from dictionary ${\bf D'}$ and it is obviously never included in ${\bf D}$. The algorithm keeps up with selecting randomly the next example ${\bf y}_i$ and iterates until all of the examples are examined and a desirable dictionary ${\bf D}$ is formed. The procedure that we have described so far is depicted in steps 1-6 of Fig. \ref{fig3}. In step 1 a random example ${\bf y}_i$ is selected and the representation errors $\mathbf{E}_{clean}$, $\mathbf{E'}_{clean}$, $\mathbf{E}_{noisy}$ and $\mathbf{E'}_{noisy}$ of stages one and two of the algorithm are computed. Fig. \ref{fig3} is a snapshot of our algorithm at some iteration $j$, as ${\bf D}$ and ${\bf D'}$ are initialized with the example ${\bf y}_4$, and the example ${\bf y}_2$ was already examined and included in dictionary ${\bf D}$, while some other examples may have also been examined but were not included in ${\bf D}$. So, at the $j^{th}$ iteration another example ${\bf y}_i$ (in blue color) is examined as to whether it will be included in ${\bf D}$ or not. Step 2 of Fig. \ref{fig3} is the combination of stages one and two of our algorithm, i.e. it is the step, where the inclusion of the example ${\bf} y_i$ in dictionary ${\bf D}$ is determined. In step 3, after we have finished with the example ${\bf y}_i$, we keep up by selecting randomly the next example ${\bf y}_{i+1}$ and steps 1-2 are repeated again for this example too. Step 3 is repeated for all the examples ${\bf} y_j's$. In step 4 we obtain the dictionary ${\bf D}$, and in step 5 we move on to the next epoch, where ${\bf D}$ will be used to initialize ${\bf D}$ and ${\bf D'}$ (step 6).

In order to report the final dictionary ${\bf D}$, the steps 1-6 of Fig. \ref{fig3} are repeated 4 times-epochs in exactly the same mode that was described previously (we use 4 epochs because as shown and discussed later in Fig. \ref{fig12}, after the third epoch the performance of the algorithm is stabilized). In every epoch of the algorithm the examples in set ${\bf Y}_{clean}$ are randomly selected and examined as to whether they will be included in the dictionary or not. Moreover, from the second epoch onward the dictionaries ${\bf D}$ and ${\bf D'}$ are not initialized with one random example as in the first epoch. Instead, the algorithm  initializes both dictionaries ${\bf D}$ and ${\bf D'}$ with the dictionary ${\bf D}$ that was formed in step 5 of the previous epoch, which is essentially used as a baseline for the construction of the next dictionaries. 

The reason for introducing the idea of epochs in our algorithm is that in every epoch new examples can be added, which in previous epochs were kept out of the dictionary, because at the time they were selected and examined some other examples with which they could make a good combination were not examined yet, and as a result at that epoch they remained out of the dictionary. After the completion of these 4 epochs the algorithm terminates and as shown in Fig. \ref{fig3} we report our final dictionary ${\bf D}$. We emphasize once more that the dictionary size does not have to be predefined by the user, as the algorithm decides itself for the number of the dictionary elements-patterns that are sufficient for the effective representation of the data. 

\subsection{Relaxed Adversarial Dictionary Learning Algorithm}
In this section we describe the RADL algorithm, which is the extension of the ADL algorithm that was described in the previous part. In addition to the synchronous activity (i.e. firing activity within the same time bin), RADL can identify temporal patterns within bigger time window intervals and outputs them to a dictionary.

\begin{figure}[!h] 
\vspace{5pt}
\centerline{\includegraphics[width=0.45\textwidth]{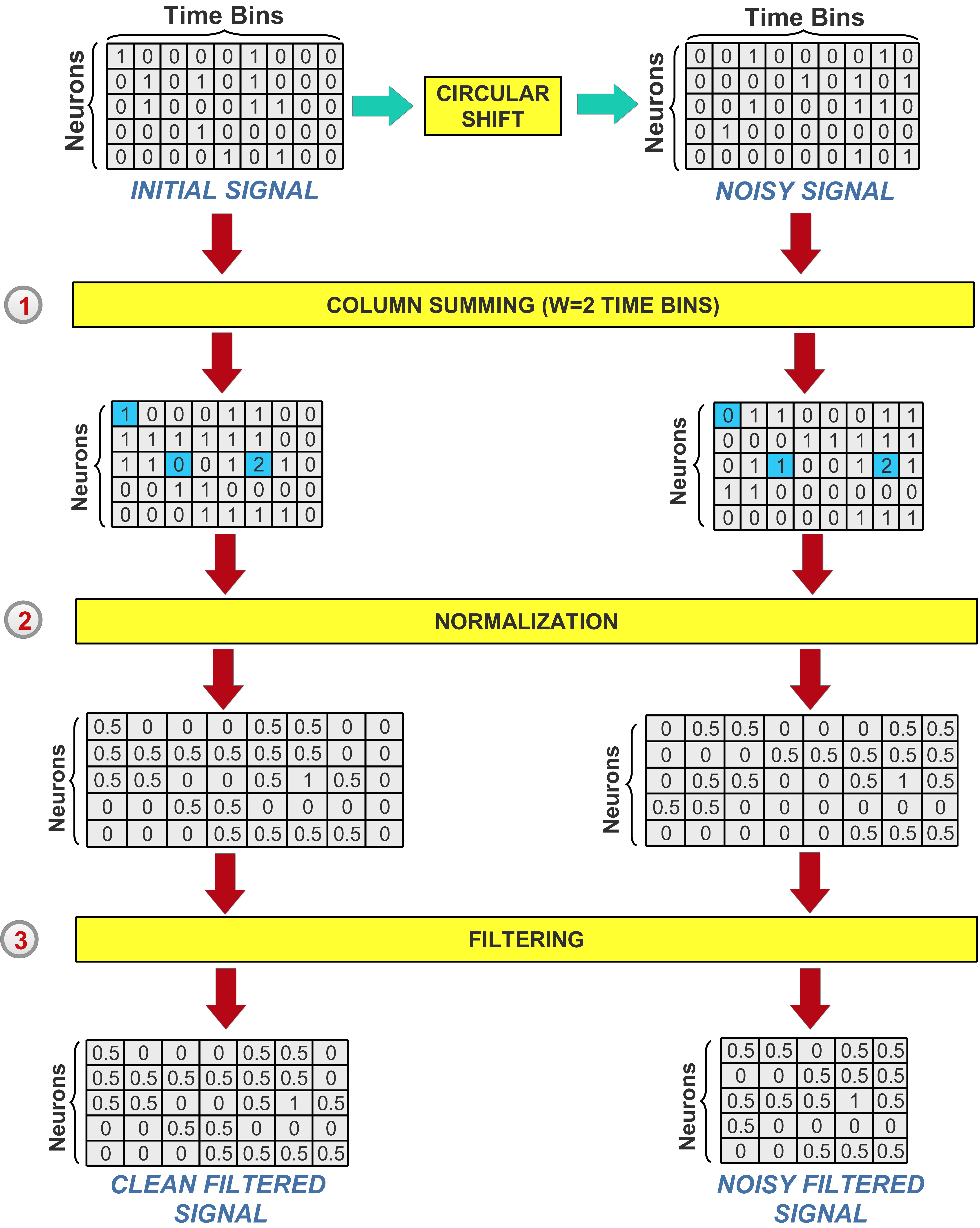}}
\caption{Searching for patterns with temporal correlation within a time window $W=2$. We sum the signal every 2 columns in an overlapping mode (step 1), we normalize the values (step 2) and we remove zero columns and those where only one neuron is active (step 3), for both initial and noisy signal.}
\label{fig4}
\end{figure}

We define a time-window parameter $W$, which determines the number of time bins that will be used, in order to search for patterns with some temporal correlation within that interval. Thus, by defining the length of the time-window to be $W$ time bins, we add the content of every $W$ columns-time bins in an overlapping mode. Namely, we sum the columns $y_1+y_2+...+y_W$,  $y_2+y_3+...+y_{W+1}$,  $y_3+y_4+...+y_{W+2}$ $etc.$ We also normalize all the values that come out from this summation by dividing with the length of the time-window (i.e. by $W$), so that the values are normalized in the scale \{0 1\}. The procedure and the idea behind this approach, i.e. the reason why the summing of the columns gives us the possibility to identify temporal patterns within bigger time window intervals is explained with the following example, which is depicted in Fig. \ref{fig4}. If we define the time window for example to be $W=2$ time bins, we add the content of every 2 columns-time bins in an overlapping mode as shown in Fig. \ref{fig4}. Namely, we sum up the columns $y_1+y_2$,  $y_2+y_3$,  $y_3+y_4$ $etc.$ and the values that come out from this summation are $0$, $1$ and $2$ (highlighted in blue). The first column of the matrix after the summations indicates that neurons 1, 2 and 3 have some temporal correlation, which is indeed true, as neurons 1, 2 and 3 in the initial signal are activated in consecutive time bins. More specifically, neuron 1 is activated exactly one time bin before neurons 2 and 3, while 2 and 3 are synchronous in the same time bin. In this mode we check temporal correlations among other neurons too. Then, at the normalization step, all values are normalized in the scale \{0 1\} by dividing with $W$ so that the and thus, values $0$, $0.5$, and $1$ for $W=2$ time bins represent: 
\begin{itemize}
\item $0$: Neuron did not fire at all within $W=2$ time bins
\item $0.5$: Neuron fired in one of the two time bins
\item $1$: Neuron fired consecutively at each time bin
\end{itemize}
Then, at the filtering step, zero columns and those with only one non-zero entry are removed. The same procedure as it is depicted in Fig. \ref{fig4} is obviously repeated for the noisy signal too. 
The summing of the columns in the initial signal results to a signal that has less zero columns and columns where only one neuron is active. We can also observe this in Fig. \ref{fig4}, where the initial signal included three zero columns and one column where only the first neuron was active, while after the summing of the columns the signal remained with only one zero column. Thus, during the filtering procedure the amount of columns that are removed is much smaller than before (i.e. when we applied the ADL algorithm and there was no column summing), which results to a training set ${\bf Y}_{clean}$ with more examples. Thus, as we increase the time window, the number of columns that have to be removed during the filtering is much smaller, which results to an increase in the number of the examples of each set as shown in Table \ref{table1}. The increase in the number of the training examples brought also an increase in the size of the dictionary, which RADL outputs and in order to compress it, apart from the sets ${\bf Y}_{clean} $, ${\bf V}_{clean}$ and the corresponding noisy sets ${\bf Y}_{noisy}$ and ${\bf V}_{noisy}$, we also introduce during the training procedure a testing set ${\bf T}_1 \in F^{M \times S}$ of $S$ clean and adversarial-noisy examples, where $F$ is the set of normalized values in scale \{0 1\}. ${\bf T}_1$ is independent from the testing set ${\bf T}_2 \in F^{M \times Q}$, where $Q$ is the number of clean and adversarial-noisy examples that will be used in the final step of the algorithm, in order to obtain the final performance of our model. 

For the compression of the dictionaries that are produced in every epoch, we use only the clean examples of the set ${\bf T}_1$ (the noisy examples of set ${\bf T}_1$ are used only after the compression to evaluate the performance of our algorithm in every single epoch). More specifically, in order to compress the dictionary formed in each epoch we remove all the dictionary elements that are not used significantly in the representation of the clean testing examples of the set ${\bf T}_1$. So, after the formation of each dictionary ${\bf D}$ (step 4 in Fig. \ref{fig3}), and before we use it in the next epoch, we examine how much each dictionary element is used for the representation of the clean examples of set ${\bf T}_1$. The contribution of each dictionary element is measured in the following way: 
Given the dictionary ${\bf D}$ that is formed in the current epoch, we obtain the Coefficient Matrix ${\bf X}$, whose columns refer to the clean testing examples of set ${\bf T}_1$ described above. For every row-vector $i$ of the Coefficient Matrix ${\bf X}$, namely for every $x_i$ that refers to the specific column-vector dictionary element $d_i$, we calculate its $l^2\textrm{-norm}$. Then, we sum all the elements of the row-vector $x_i$ and if the summation is smaller or equal with the $l^2\textrm{-norm}$, then we remove the element $d_i$ from the dictionary. The intuition behind this technique is that we remove all dictionary elements that are used negatively for the representation of most of the examples (i.e. when row-vector $x_i$ has many negative values). Eventually, in the last epoch of the algorithm (i.e. the $4^{th}$ epoch) we obtain the final dictionary ${\bf D}$, which is used with the testing set ${\bf T}_2$ that we have available for the testing procedure, in order to evaluate the performance of our algorithm.

So, what essentially changes from ADL is the input data that we give to the system, where every column-time bin in the new data represents patterns that have a temporal correlation within $W$ time bins. Obviously, this information but in a compressed format is also encoded in the dictionary, providing an insight into temporal correlations. Additionally, during the training procedure of the RADL algorithm, we compress the dictionary of each epoch by removing the dictionary elements that have small contribution in the representation of the clean examples in ${\bf T}_1$.

\subsection{Evaluation of the dictionary quality}
\noindent In order to evaluate the quality of the output dictionaries in terms of learning capacity and robustness to noise, we employ a supervised machine learning framework by training an SVM-classifier with the clean and noisy raw data as well as with the reconstructed ones (i.e. the output of $DX$).  We aim to examine the extent to which the classifier can discriminate the clean from the noisy activation patterns, and whether its training with the reconstructed data results to a better classification performance, rather than when we use the raw data. Thus, the classification performance is the quantitative metric offering an insight about the extent to which the output dictionary has captured the underlying statistics of the data.

\section{Performance Analysis}

\subsection{Dataset Collection}
\noindent To evalute the merits of the proposed modeling approach, we employed two real-world datasets that were collected using two-photon calcium imaging in the neocortex of a 9-day old mouse and a 36-day old one (C57BL/6). The first dataset of the 9-day old mouse includes 183 neurons of the layer 2/3 of the V1, and neurons were imaged using calcium indicator OGB-1 (imaging depth 130 microns from pia). The dataset of the 36-day old mouse includes 126 neurons of the layer 2/3 of the V1 area. Additionaly, for the 9-day old mouse 29 minutes of spontaneous activity were recorded, comprised of 11970 frames, each of 0.1451 seconds duration, while for the older one the total movie length was 30 minutes comprised of 11972 frames, each of 0.15 seconds duration. The raw fluorescence movie was motion-corrected to remove slow xy-plane drift. After motion correction, we used ImageJ software \cite{abramoff2004image} to draw the ROIs of cells around cell body centers, staying 1-2 pixels from the margin of a cell in the case of the 9-day old mouse, in order to avoid contamination with neuropil signals and 1-2 pixels for the 36-day old mouse. We then averaged the signals of cell ROI pixels and converted them to dF/F \cite{stosiek2003vivo}. To determine the onsets of spontaneous calcium responses, the dF/F timecourse for each cell was thresholded, using the noise portion of the data, to 3 standard deviations above noise. To make a binary eventogram of the responses, for each cell the frames containing the onsets for this particular cell were assigned the value 1, and all other frames were assigned the value 0. The resulting binary eventogram of all cells was used in subsequent analysis.

\subsection{Adversarial Dictionary Learning (ADL)}
In this section, we illustrate the performance of our proposed algorithm ADL for the case of one time bin window interval ($W=1$), with respect to other methods, such as K-SVD \cite{aharon2006rm}, Analysis K-SVD \cite{rubinstein2012analysis}, LC-KSVD \cite{jiang2013label} and ODL \cite{mairal2009online}. More specifically, we examine which of the trained dictionaries produced from these methods are more robust to adversarial noise. In order to quantify this information, we examine the extent to which each trained dictionary can discriminate the clean from the adversarial-noisy activation patterns. Through this analysis the impact of the following parameters is also explored:
\begin{itemize}
\item Dictionary size ($DS$), which is the number of elements considered in the dictionary. While in all examined methods, $DS$ must be defined by the user, in our method, it is automatically inferred.
\item Sparsity level ($SL$), i.e., the maximal number of dictionary elements that are used for representation of the examples.
\end{itemize}
We also present some more qualitative results of the dictionary that are produced from our proposed method.

\subsubsection{Parameter Setup}
After the completion of the filtering that is described in Fig. \ref{fig2}, we select $50\%$ of the examples of the clean filtered signal, namely 1138 examples to train K-SVD. Regarding our proposed method, in order to train the dictionary we select the same $50\%$ examples from the clean filtered signal, as well as $50\%$ of the examples from the noisy filtered signal. Subsequently, the other half of the clean and noisy filtered signal sets will serve as the testing set for each one of the two methods. Namely, they will be used for the training and testing of an SVM-classifier with gaussian kernel and scale $\sigma=0.01$. The classifier is trained and tested with the:
\begin{enumerate}
\item [(i)] Raw clean and noisy data
\item [(ii)]Reconstructed clean and noisy data, which are binarized by considering 
all values greater than $0.5$ as activations ($1s$), while the rest as zeros.
\end{enumerate} 
The number of the testing examples in set ${\bf T}_2$ as well as the number of the training examples in set ${\bf Y}$, where ${\bf Y}$ consists of the clean examples ${\bf Y}_{clean}$ and the adversarial-noisy examples ${\bf Y}_{noisy}$ for the case of one time bin window interval ($W=1$) are depicted in Table \ref{table1}. Note that all sets described in Table \ref{table1} (${\bf Y}$, ${\bf T}_1$ and ${\bf T}_2$) include the number of both the clean and the adversarial-noisy examples (i.e. half of the size of each set described in Table \ref{table1} refers to the clean examples and the other half refers to the adversarial-noisy examples).

\begin{figure}[!h]
\centering
\subfloat[]{\includegraphics[width=0.25\textwidth]{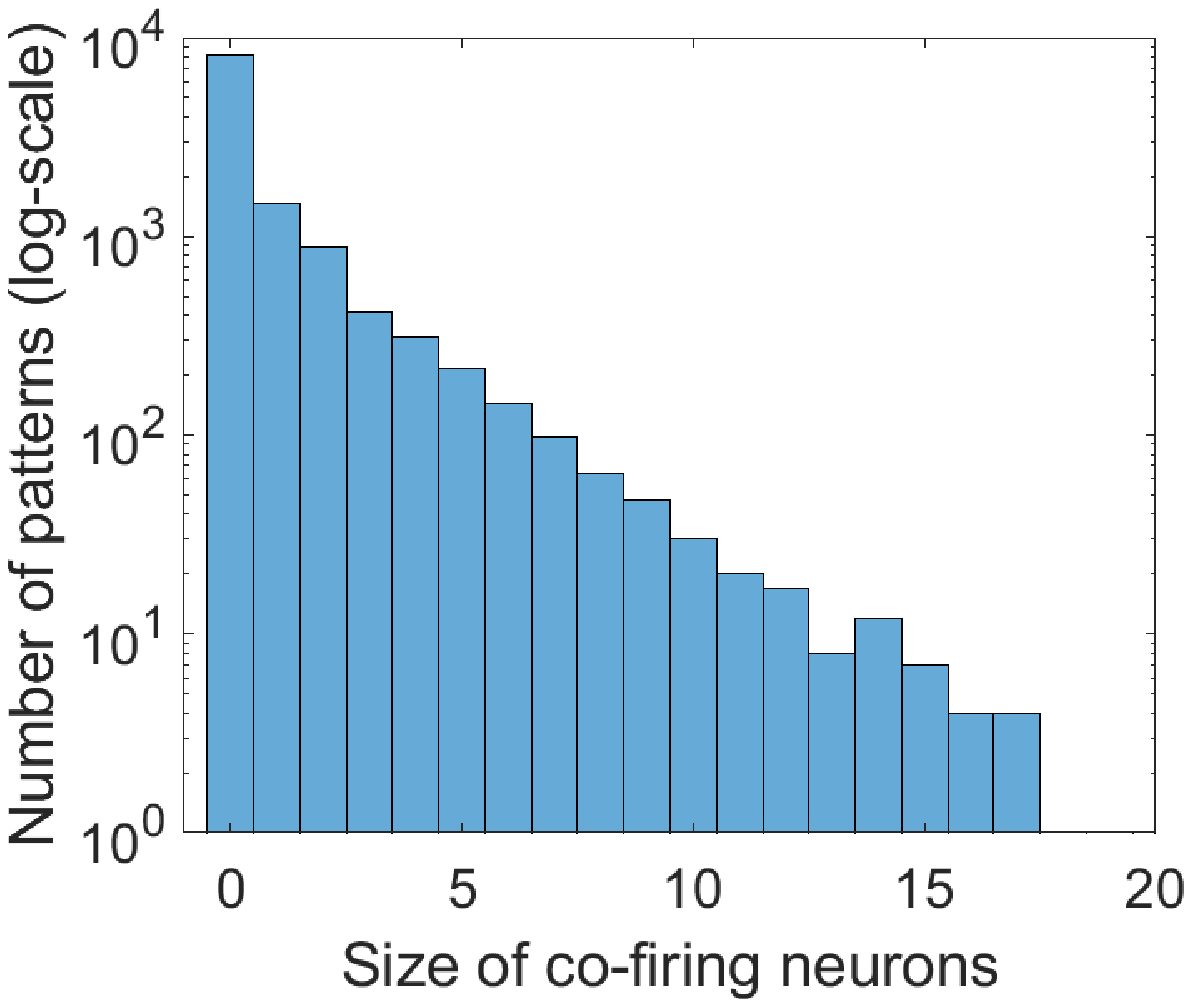}}
\subfloat[]{\includegraphics[width=0.25\textwidth]{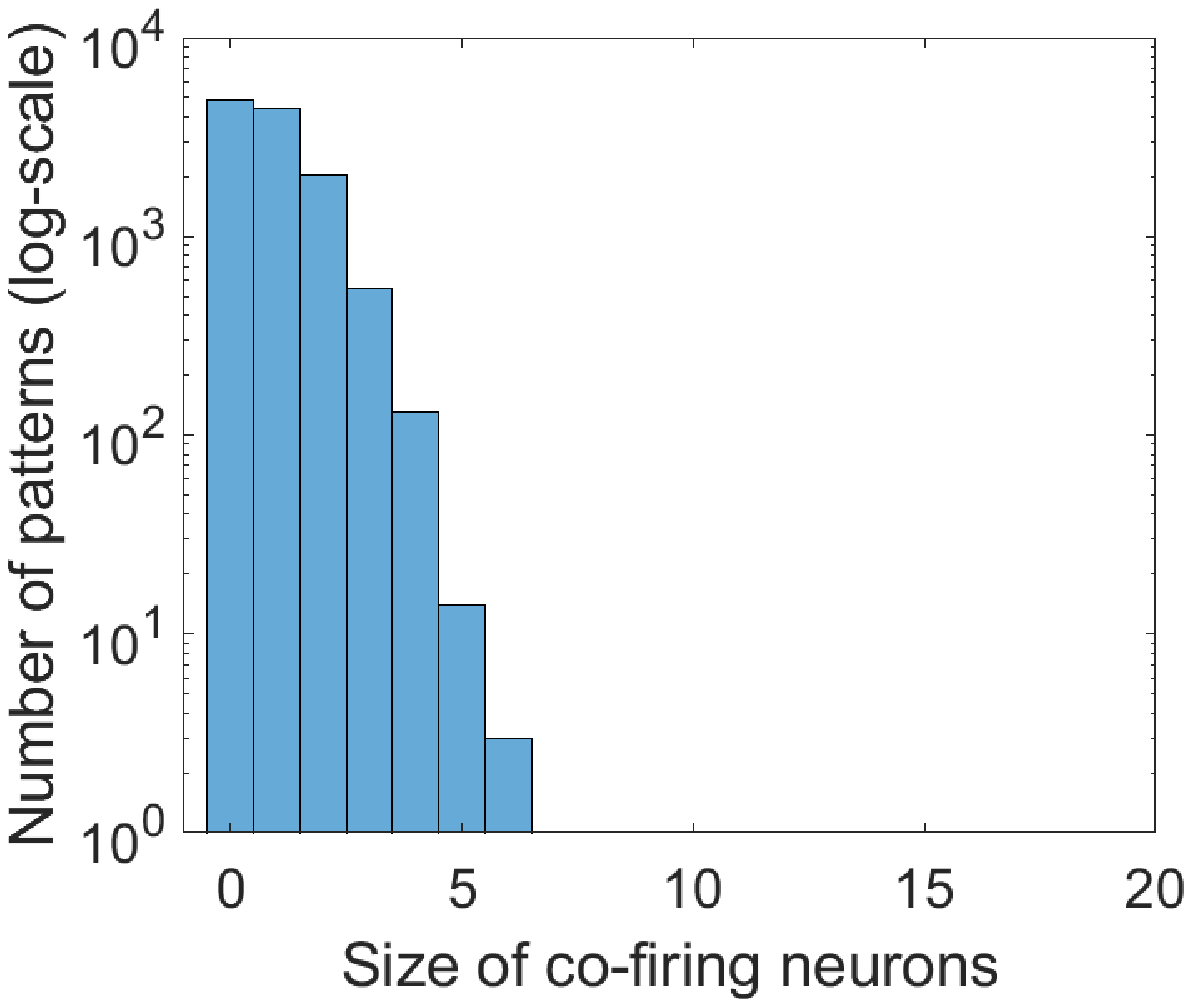}}
\caption{(a) Clean signal distribution (b) Noisy signal distribution }
\label{fig5}
\end{figure}

Fig. \ref{fig5} shows the distribution of the original clean (\ref{fig5} (a)) and of the noisy signal (\ref{fig5} (b)), as it results from the circular shifting procedure. The distributions refer to the activity of the 9-day old mouse before the process of the filtering. Namely, in both figures axis x indicates the size of co-firing neurons (i.e. the number of neurons that co-activate within one time bin) and the log-scaled axis y indicates the number of these patterns that exist in the data. We observe that for the noisy signal, circular shifting has caused a reduction in zero columns-patterns and a simultaneous increase in doublets (i.e. patterns where 2 neurons co-activate within a time bin) as well as in patterns where only one neuron is active within a time bin. Finally, more complex patterns with more than seven neurons firing simultaneously are completely destroyed.

\begin{table}[]
\centering
\begin{tabular}{|c||c|c|c|c|}
\hline
Size             		& $W=1$ & $W=2$ & $W=3$ & $W=4$ \\ \hline \hline
Training Set $(Y)$ 		& 2276   & 2964   &  3648  & 4156   \\ \hline
Testing Set $(T_{1})$ 	&  -  & 1866   & 2270   & 2578  \\ \hline
Testing Set $(T_{2})$ 	& 2324   & 2744   &  3336  &  3770  \\ \hline
\end{tabular}
\vspace{5pt}
\caption{Sizes of the Sets $Y$, $T_1$ and $T_2$ for all $Ws$}
\label{table1}
\end{table}

\subsubsection{Evaluation Results} 
\begin{figure}[!h] 
\centerline{\includegraphics[width=0.27\textwidth]{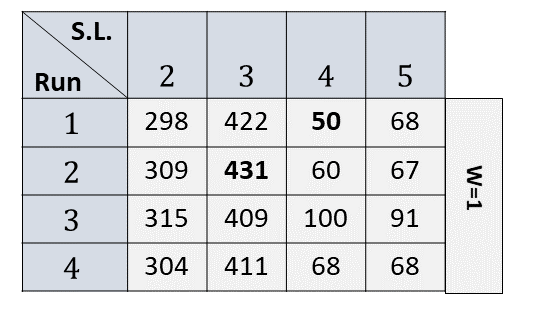}}
\caption{Size of the final dictionary D for every run and Sparsity Level (SL).}
\label{fig6}
\end{figure}

\begin{figure*}[!h]
\centering
\subfloat[DS=150]{\includegraphics[width=0.32\textwidth]{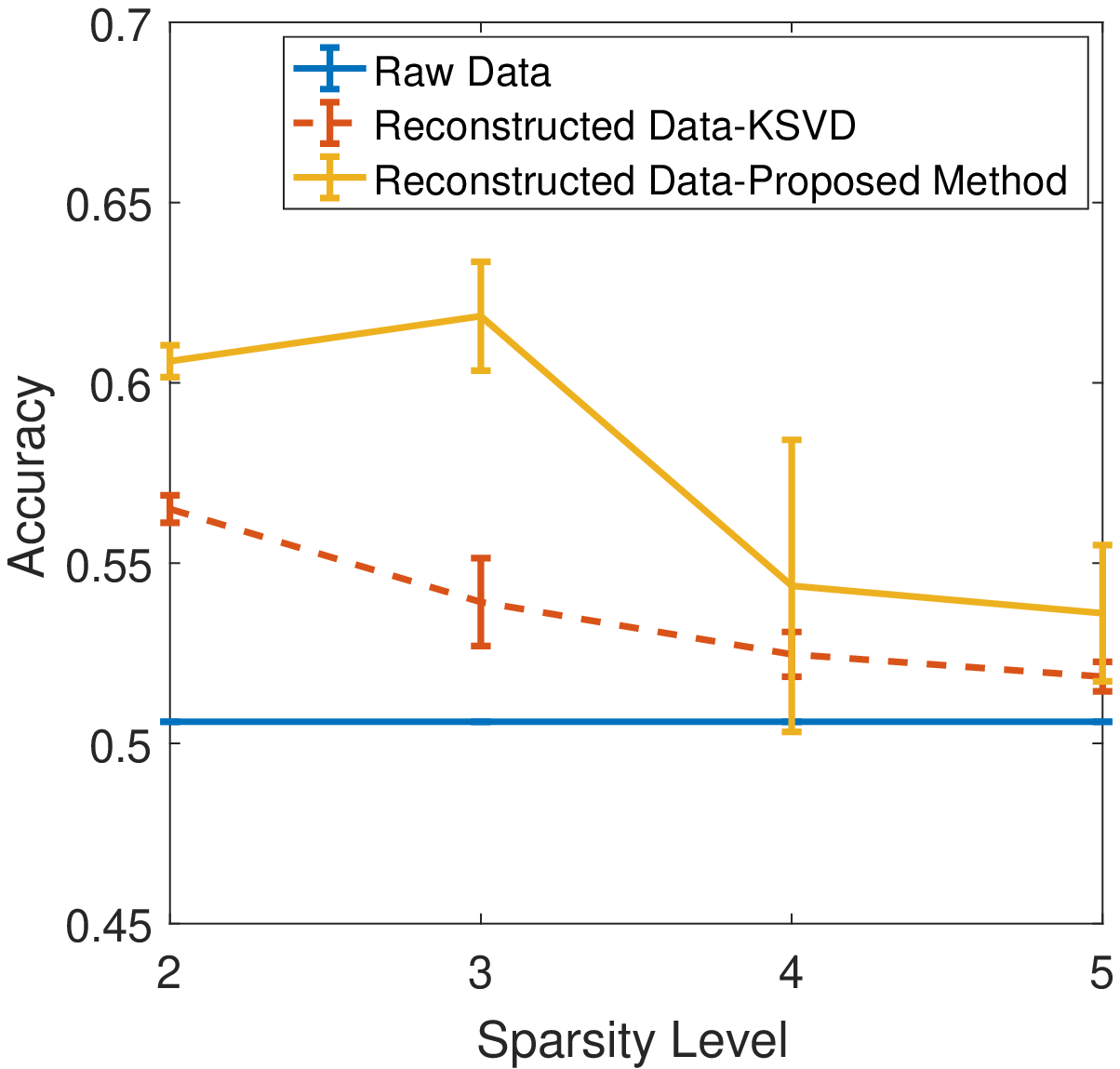}}
\subfloat[DS=200]{\includegraphics[width=0.32\textwidth]{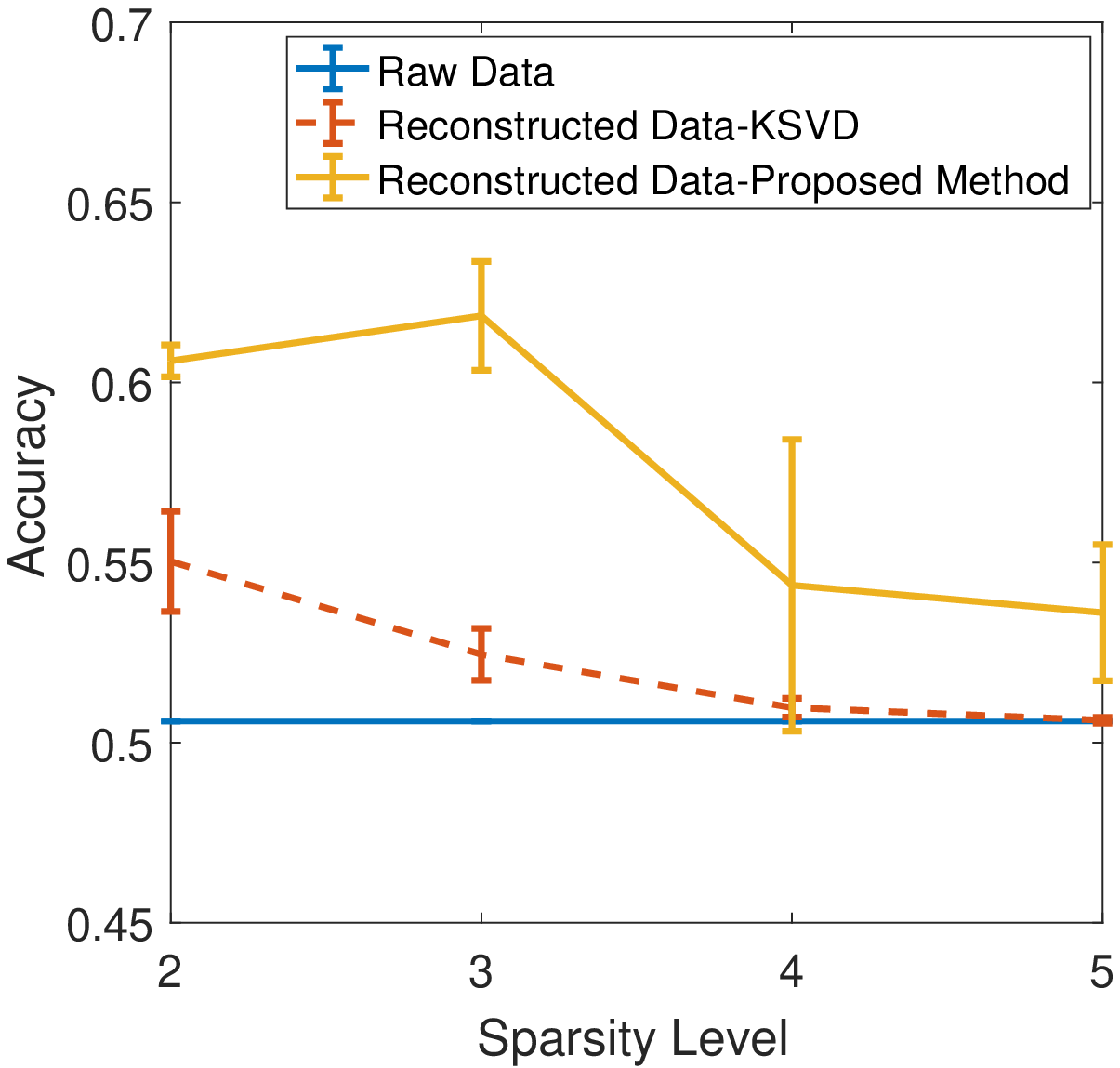}}
\subfloat[DS=300]{\includegraphics[width=0.32\textwidth]{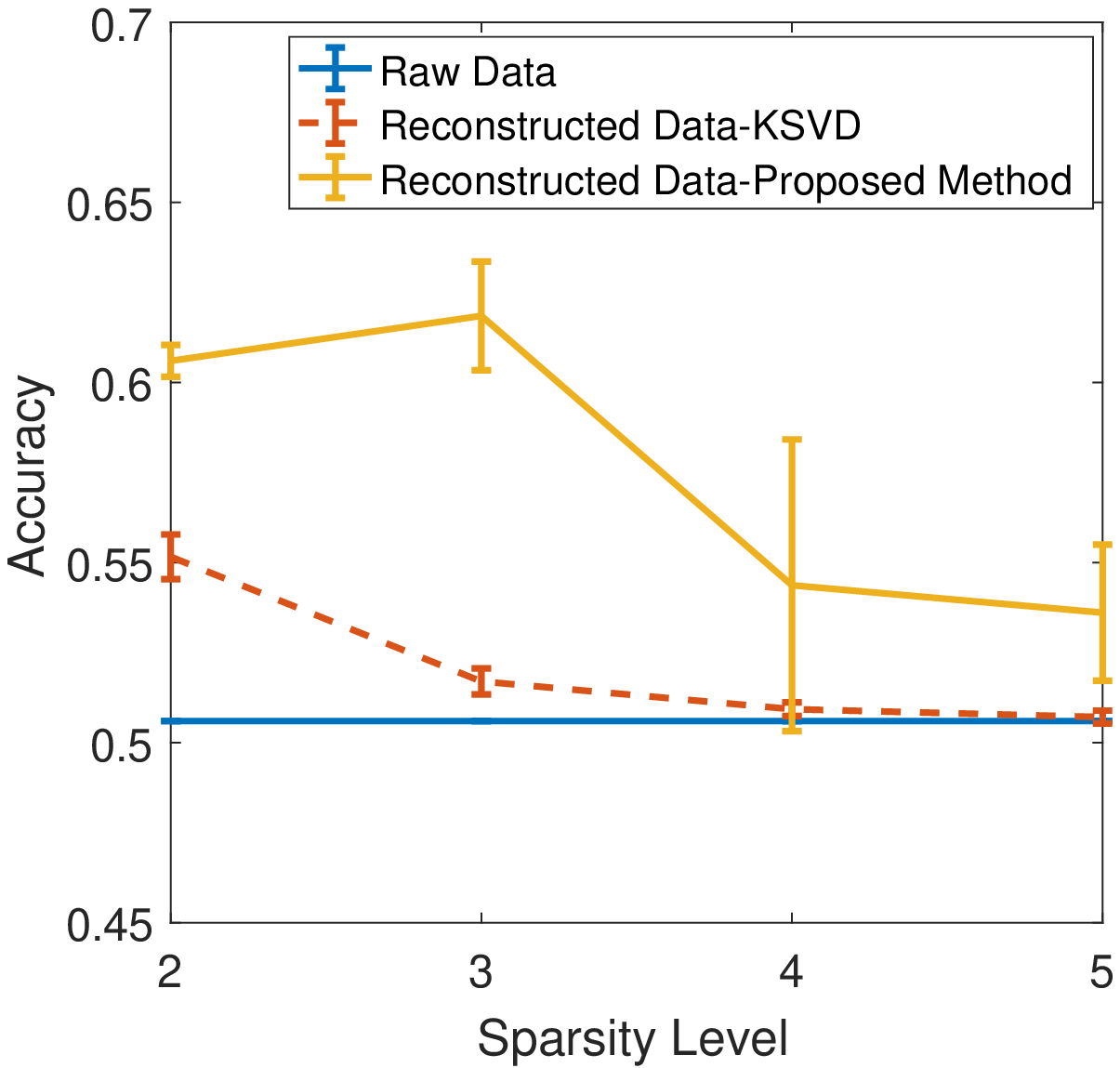}}
\caption{Mean accuracy classification performance when the classifier is trained with the raw data, the reconstructed data produced by our method ADL and the reconstructed data produced by the K-SVD.}
\label{fig7}
\end{figure*}

Fig. \ref{fig7} illustrates the performance of the SVM-classifier regarding the discrimination between the clean and the noisy signals for the 9-day old mouse, as a function of the sparsity level when the classifier is trained and tested with the raw data, the reconstructed data produced by our proposed method ADL and the reconstructed data produced by the K-SVD algorithm. Each point in the errorbar plots corresponds to the mean accuracy of 4 runs and in every run the examples in the training set are given with a different sequence in terms of the columns (i.e the second column of the training set in the first run may be the fifth column of the training set in the second run). These 4 runs are executed in order to examine the sensitivity of our algorithm with respect to the different sequence that the examples are selected. Thus, the K-SVD algorithm is initialized with a different dictionary in every run, as the columns are presented with a different sequence. Regarding our algorithm, the different sequence in the columns of the training set in every run, results to the selection and as a consequence to the examination of the examples with a different sequence as to whether they will be included in the dictionary $D$ or not. The testing set remains the same in all runs. The vertical error bar demonstrates the standard deviation of these four runs (i.e. how much the accuracy of each run differs from the mean accuracy of the four runs). More specifically, as it is illustrated in each subfigure of Fig. \ref{fig7}, we give as input to the K-SVD algorithm a different dictionary size, and we evaluate the performance of the algorithm compared to our proposed method. Fig. \ref{fig6} depicts the corresponding dictionary sizes that are produced from our method for the case of $W=1$. More specifically, for every sparsity level (SL), Fig. \ref{fig6} demonstrates the size of the final dictionary $D$ that is obtained from the $4^{th}$ epoch for each one of the 4 runs.\\

We observe in Fig. \ref{fig7}  that when the classifier is trained and tested with the raw data, the accuracy that it achieves is almost $51\%$. This percentage is quite low and indicates the difficulty of the problem that we are supposed to solve. By using the reconstructed data that are produced by the K-SVD algorithm we observe that the classifier achieves a better performance with an accuracy of $56\%$ for DS=150 elements and for SL=2. In all of the subfigures we observe that as the SL increases, the accuracy of the classifier decreases, which can be attributed to overfitting of the system. Moreover, the three different dictionary sizes, which were tried as input to the K-SVD algorithm do not affect significantly the performance of the classifier. When we use the reconstructed data that are produced from our method and as depicted in Fig. \ref{fig7}, the classifier achieves better performance results compared to the performance of the K-SVD algorithm. More specifically, we obtain an accuracy of $62\%$ for SL=3 and mean dictionary size (of the 4 runs) equal to 418. We observe that for values of sparsity level greater than 3 the performance deteriorates due to overfitting. Nevertheless, our proposed method gives better performance results for every value of sparsity level. 

In Table \ref{table2}, we report the mean accuracy performance of 4 runs for several DL methods and for various values of DS and SL. The parameters that we used for each method were selected after exhaustive search, so that they give the best possible accuracy performance. Regarding ODL and Analysis K-SVD, the SL parameter is only used with the OMP algorithm to obtain the coefficient matrix corresponding to the testing data (for their reconstruction) and not during the training procedure (i.e to obtain the output dictionary). We observe that Analysis K-SVD outperforms all the other methods for all the examined parameters, but still gives a worse accuracy performance compared to ADL. The corresponding DSs of ADL for each SL are reported in Fig. \ref{fig11}.

We also applied the PCA method, which is a dimensionality reduction algorithm, not dictionary learning based, on both the clean and the noisy test data. We obtained the corresponding coefficients and used them in order to train and test the classifier, which gave an accuracy performance of $51.55\%$

\begin{table}[h]
\scriptsize
\renewcommand{\arraystretch}{1.4}
\begin{tabular}{|c|c|c|c|c|c|c|c|}
\hhline{|=|=|=|=|=|=|}
\textbf{Methods} & \textbf{DS} & \textbf{SL=2} & \textbf{SL=3} & \textbf{SL=4} & \textbf{SL=5} \\ \hhline{|=|=|=|=|=|=|}
\multirow{3}{*}{ODL} & \begin{tabular}[c]{@{}c@{}}150\end{tabular} & $0.508$ & $0.5161$ & $0.5157$ & $0.5155$ \\ \cline{2-8} 
& 200 & $0.5105$ & $0.5062$ & $0.5065$ & $0.5065$ \\ \cline{2-8} 
& 300 & $0.5077$ & $0.5056$ & $0.506$ & $0.506$ 
\\ \hhline{|=|=|=|=|=|=|}
\multirow{3}{*}{LC-KSVD1} & \begin{tabular}[c]{@{}c@{}}150\end{tabular} & $0.5077$ & $0.4976$ & $0.5063$ & $0.4996$ \\ \cline{2-8} 
& 200 & $0.5041$ & $0.5011$ & $0.5006$ & $0.5025$ \\ \cline{2-8} 
& 300 & $0.5069$ & $0.5032$ & $0.5018$ & $0.5037$ 
\\ \hhline{|=|=|=|=|=|=|}
\multirow{3}{*}{LC-KSVD2}   &  \begin{tabular}[c]{@{}c@{}} 150 \end{tabular} & $0.5267$ & $0.5077$ & $0.5188$ & $0.51$\\ \cline{2-8} 
& 200 & $0.5284$ & $0.542$ & $0.5388$ & $0.5267$ \\ \cline{2-8} & 300 & $0.5297$ & $0.5374$ & $0.5138$ & $0.5211$\\ \hhline{|=|=|=|=|=|=|} 
\multirow{3}{*}{Analysis K-SVD}   &  \begin{tabular}[c]{@{}c@{}} 150 \end{tabular} & $0.5553$ & $0.5577$ & $0.5639$ & $0.5678$\\ \cline{2-8} 
& 200 & $0.5688$ & $0.5749$ & $0.5747$ & $0.5818$ \\ \cline{2-8} & 300 & $0.5658$ & $0.5617$ & $0.559$ & $0.5663$\\ \hhline{|=|=|=|=|=|=|} 
\multirow{1}{*}{ADL}   &  \begin{tabular}[c]{@{}c@{}} - \end{tabular} & $0.6059$ & $\bf{0.6185}$ & $0.5436$ & $0.5436$ 
\\ \hhline{|=|=|=|=|=|=|}
\end{tabular}\\[4pt]
\caption{Mean accuracy performance when the classifier is trained with the reconstructed data produced by ODL, LC-KSVD1, LC-KSVD2, Analysis K-SVD and ADL.}
\label{table2}
\end{table}

As it is already stated, our algorithm executes 4 runs, where in every run the examples of the training set are selected and examined with a different sequence as to whether they will be included in the dictionary or not. 
\begin{figure}[!h]
\centerline{\includegraphics[width=0.35\textwidth]{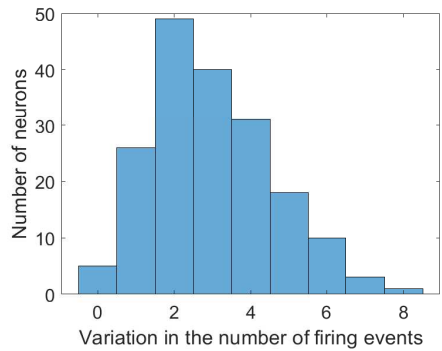}}
\caption{Neurons grouped in the same bin have the same variation in the number of firing events across the 4 dictionaries formed in every run ($W=1$, Sparsity Level=2).}
\label{fig14}
\end{figure}
Thus, we want to ensure that neurons' firing activity captured by the dictionaries of each run will be similar and not with intense variations. To that end, we demonstrate Fig. \ref{fig14}, which depicts the variation in the number of firing events that neurons have across the 4 dictionaries formed in each run, under the consideration of $W=1$ and $SL=2$.
We observe that for most of the neurons (almost 50 neurons) the maximum variation across the 4 dictionaries is only 2 firing events, while only one neuron has a variation of 8 firing events. Thus, we end up with 4 dictionaries that have almost the same number of firing events for each neuron, indicating the robustness of our algorithm with respect to the different sequence in the selection of the examples.

Unlike all the methods that we compared, which produce real-numbered dictionaries with no physical meaning for our application, our proposed method ADL produces dictionaries that provides us with quantitative as well as with qualitative information, giving us an insight about the synchronicity patterns existing in the data. 
\begin{figure}[!h]
\centering
\centerline{\includegraphics[width=0.35\textwidth]{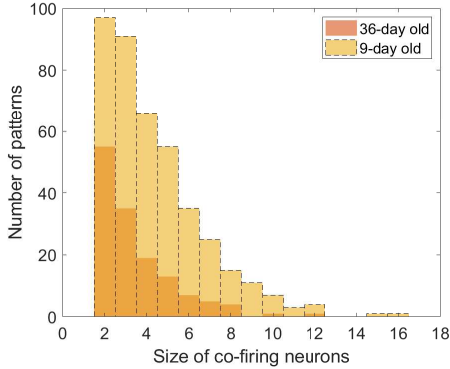}}
\caption{Distribution of the two dictionaries ($W$=1, Sparsity Level=3).}
\label{fig8}
\end{figure}
So, Fig. \ref{fig8} demonstrates the distribution of two dictionaries (we used the dictionaries that were produced from the $4^{th}$ run of our algorithm, for $SL=3$) that refer to the spontaneous neuronal activity of a 9-day old and a 36-day old mouse. Namely, axis x indicates the size of the co-firing neurons that exist in the dictionary, i.e. the number of neurons that co-activate within one time bin, such as doublets (when 2 neurons co-activate within one time bin) or triplets (when 3 neurons co-activate within one time bin), etc and axis y indicates the number of these patterns (doublets etc.) that exist in the dictionary. The dataset that refers to the 9-day old mouse, firing events occupy the $0.487\%$ of the data, while for the 36-day old mouse firing activity occupies only the $0.364\%$ of the dataset. These percentages show the sparseness of our datasets and by extension indicate the low frequency of the neurons' firing activity for both laboratory animals. Moreover, these percentages reveal that the 9-day old mouse has a more intense firing activity, which can be attributed to its young age. All this information is depicted in the distribution of the two trained dictionaries Fig. \ref{fig8}, as we observe that the number of the various synchronicity patterns for the 9-day old mouse is greater than the number of patterns for the 36-day old mouse. Additionally, the dictionary that refers to the activity of the 9-day old mouse includes more complex patterns with more than six neurons firing simultaneously, while for the 36-day old mouse such patterns tend to be zero. Eventually, the size of each dictionary also reveals information about the data that we summarize. Namely, the dictionary that refers to the activity of the 9-day old mouse has a size of 411 elements as depicted in Fig. \ref{fig6}, while the dictionary that refers to the older mouse has a size of 51 dictionary elements, which correctly verifies that it fires less.

\subsection{Relaxed Adversarial Dictionary Learning (RADL)}
This section demonstrates the analysis for temporal correlation patterns within larger time window intervals $(W>1)$. The analysis assesses the impact of the following parameters:
\begin{itemize}
\item Time window interval ($W$), from which we can extract information about temporal correlations.  
\item Sparsity level ($SL$), i.e., the maximal number of dictionary elements that are used for representation.
\end{itemize} 

\subsubsection{Parameter Setup}
\begin{figure}[!h] 
\centerline{\includegraphics[width=0.3\textwidth]{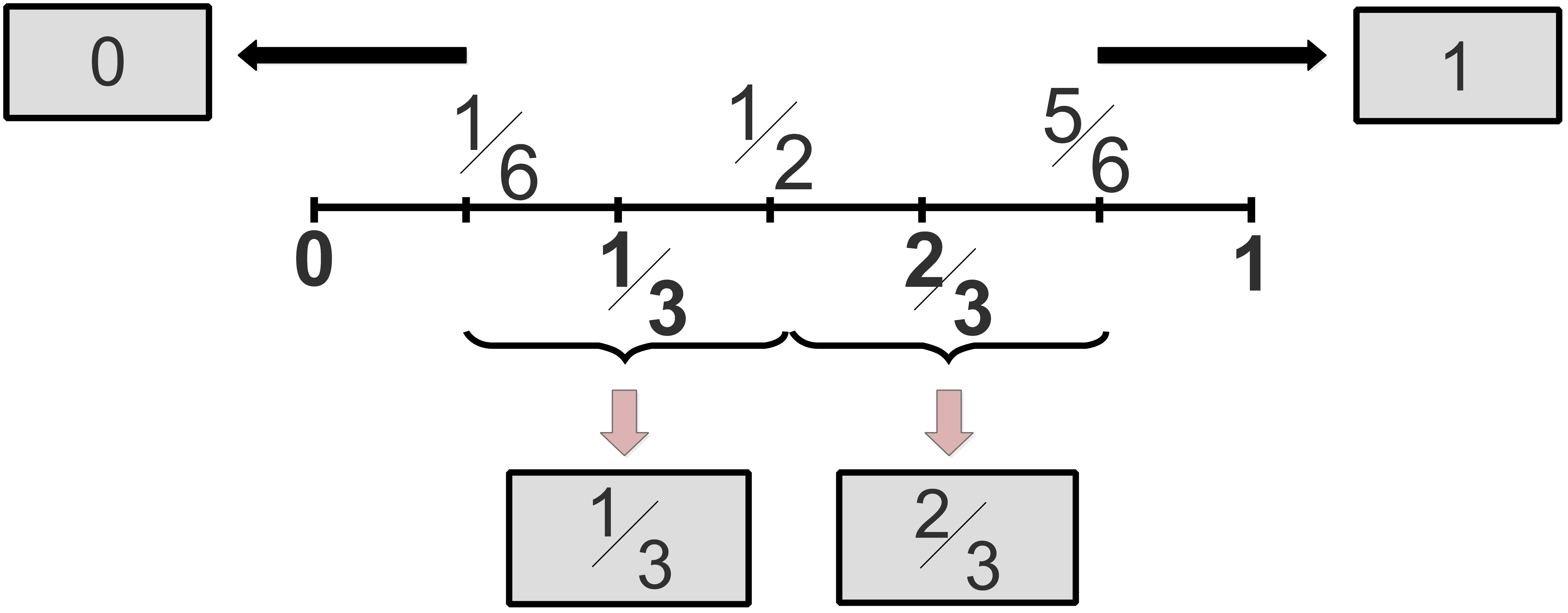}}
\caption{Processing the values of the reconstructed events.}
\label{fig9}
\end{figure}

After the completion of the procedure that is described in Fig. \ref{fig4} we select $40\%$ of the examples of the clean filtered signal, as well as $40\%$ of the examples of the noisy filtered signal for the set ${\bf Y}$, which will be used for the training of the dictionary. Then, we select $25\%$ of the examples of the clean filtered signal for the set ${\bf T}_1$, which will be used for the compression of the dictionaries that are produced in every epoch as well as $25\%$ of the examples of the noisy filtered signal in order to evaluate the performance of our algorithm at every epoch of each run. Eventually, the other $35\%$ of the clean and noisy filtered examples will be used by the set ${\bf T}_2$ and will serve as the testing set, whose half of the examples will be used for the training of an SVM-classifier with gaussian kernel and scale $\sigma=0.01$ and the other half will be used for the testing of the classifier. The number of the training examples in set ${\bf Y}$, as well as the number of the testing examples in sets ${\bf T_1}$ and ${\bf T}_2$ for all the time window intervals are depicted in Table \ref{table1}. As it was also stated in the parameter setup section of ADL, all sets described in Table \ref{table1} (${\bf Y}$, ${\bf T}_1$ and ${\bf T}_2$) include the number of the clean and the adversarial-noisy examples. The classifier is trained and tested with the:

\begin{enumerate}
\item [(i)] Raw clean and noisy data
\item [(ii)]Reconstructed clean and noisy data whose values are processed as we describe in the following example
\end{enumerate} ~\\[2pt]
As it was described in section II, for the cases of time window intervals, where $W>1$, activation patterns are not represented by the values $0$ and $1$ due to the summing of the columns and the normalization step. For example in the case of $W=3$, if one neuron has not fired at all within 3 consecutive time bins, we get a 0-event. If it has fired once, we obtain the normalized value of \( \frac{1}{3} \), which are the most prevalent values with the 0 value. Additionally, if the neuron has fired twice, we obtain the value \( \frac{2}{3} \) and if it has fired consecutively in all of the 3 time bins, we obtain a 1-event, which is not very common due to the refractory period. Because of the fact that we deal with a reconstruction problem, reconstructed values other than those described before may appear. Thus, without loss of generality we make the simplification, which is depicted in Fig. \ref{fig9}. Namely, for $W=3$ all values which are smaller than \( \frac{1}{6} \) are turned into zero. Values in space $\left[ \frac16,  \frac12 \right)$ are turned into \( \frac{1}{3} \) and values in space $\left[ \frac12,  \frac56 \right)$ are turned into \( \frac{2}{3} \). Any other value is turned into 1. Accordingly, we work for any time window $W$.

\subsubsection{Evaluation Results}
\begin{figure}[!h]
\centering
\subfloat[W=1,W=2]{\includegraphics[width=0.29\textwidth]{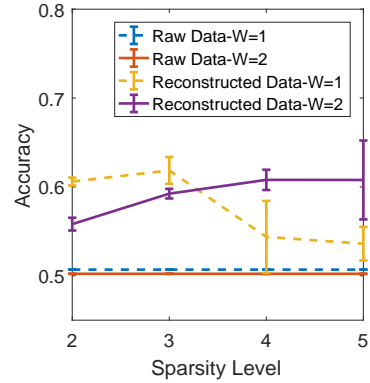}}\\
\subfloat[W=3,W=4]{\includegraphics[width=0.29\textwidth]{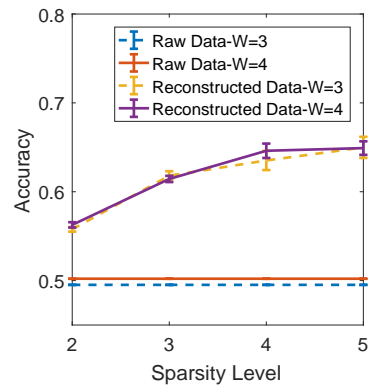}}
\caption{Classification performance when the classifier is trained with the raw data and the reconstructed data produced by RADL with respect to different time window intervals.}
\label{fig10}
\end{figure}

\begin{figure}[!ht] 
\vspace{5pt}
\centerline{\includegraphics[width=0.45\textwidth]{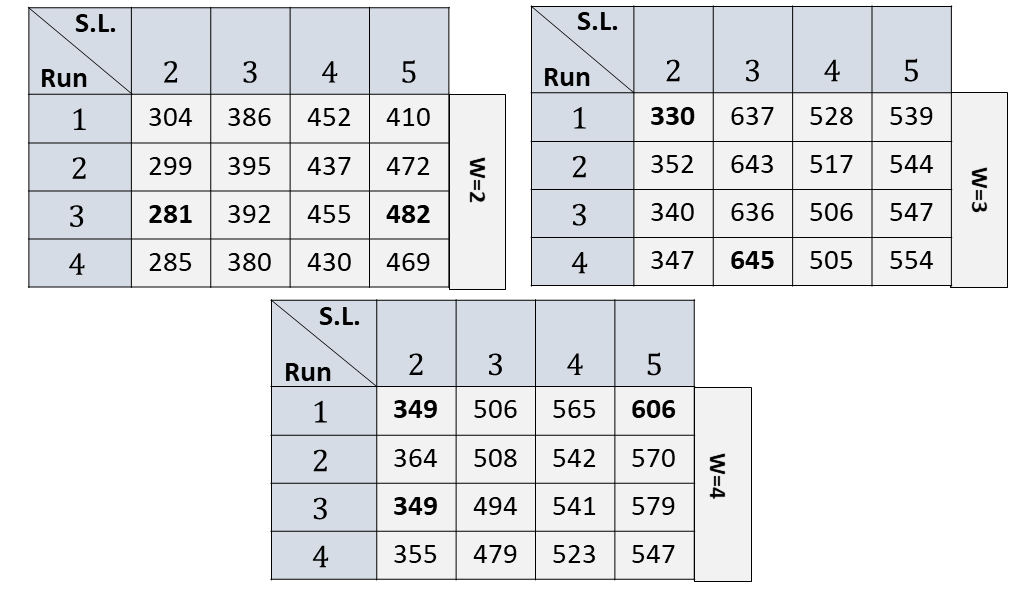}}
\caption{Size of the final dictionary D for every run and Sparsity Level (SL)}
\label{fig11}
\end{figure}

Fig. \ref{fig10} illustrates the performance of the SVM-classifier regarding the discrimination between the clean and the noisy signals for the 9-day old mouse, as a function of the SL when the classifier is trained and tested with the raw data and the reconstructed data produced by the RADL algorithm. Each point in the errorbar plots corresponds to the mean performance of the 4 runs of the algorithm, where in every run the examples in the training set are selected and examined with a different sequence as to whether they will be included in the dictionary $D$ or not. The vertical error bar demonstrates the standard deviation of these 4 runs. More specifically, as it is illustrated in Fig. \ref{fig10}, each subfigure refers to the performance of the classifier for different time window intervals.
When the classifier is trained and tested with the raw data, the highest accuracy that it achieves, taking into account all the time windows is $51\%$, which is a quite low percentage. When we use the reconstructed data that are produced from our proposed method, we observe that as we increase the time window interval, we obtain a better classification performance. More specifically, for $SL=5$ and $W=3$ as well as $W=4$, we obtain the highest accuracy performance equal to $65\%$. Moreover, we notice that for time window intervals $W>1$, when the SL is increased, the classification performance is increased too. This happens because the patterns for time windows $W>1$ are greater in number (Table \ref{table1}) and more complex (more firing events per signal) compared to the patterns of $W=1$. Thus, by increasing the SL, the algorithm obtains greater flexibility, as it can use more dictionary elements to represent the data. Consequently, the algorithm can better generalize and does not overfit as in the case of $W=1$. 

Fig. \ref{fig12} illustrates the classification performance that is obtained in every epoch of the algorithm for all the runs and for SL=3. We observe that for all the cases of time windows the classification performance is either improved or it remains the same in every epoch of the algorithm. Thus, as it is depicted in Fig. \ref{fig12} the dictionary that is obtained in the $4^{th}$ epoch of each run, ensures the best possible accuracy performance for the specific run compared to the dictionaries that are formed in the previous epochs. 

\begin{figure}
\centering
\centerline{\includegraphics[width=0.35\textwidth]{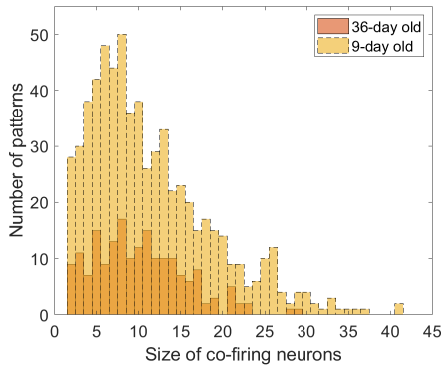}}
\caption{Distribution of the two dictionaries ($W$=3, Sparsity Level=3).}
\label{fig13}
\end{figure}

\begin{figure*}
\centering
\subfloat[W=1]{\includegraphics[width=0.25\textwidth]{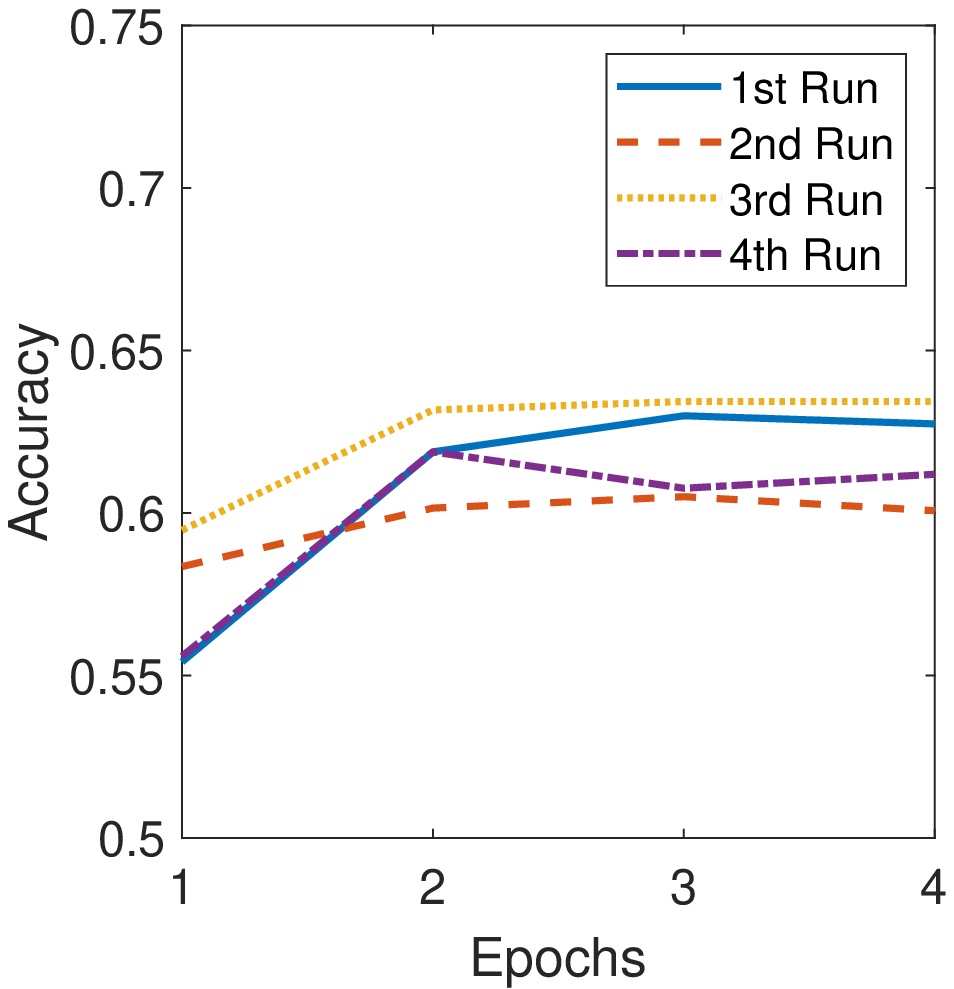}}
\subfloat[W=2]{\includegraphics[width=0.25\textwidth]{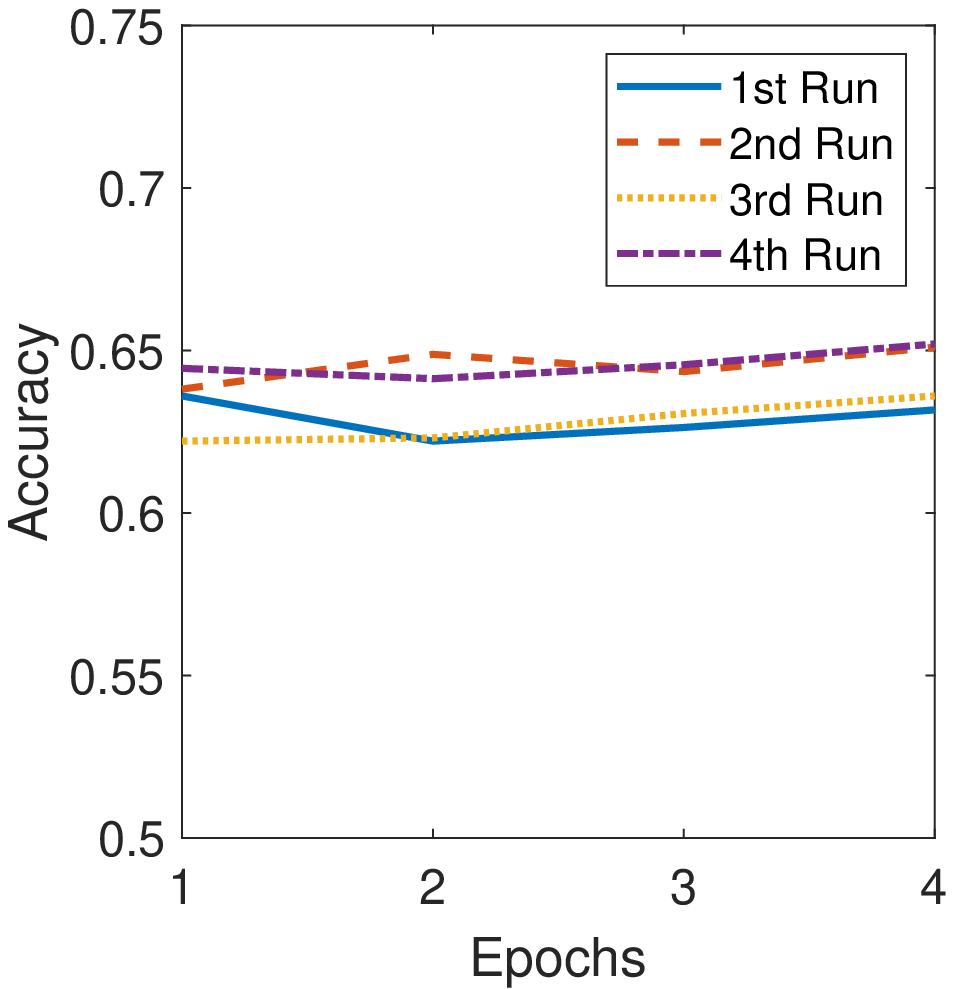}}
\subfloat[W=3]{\includegraphics[width=0.25\textwidth]{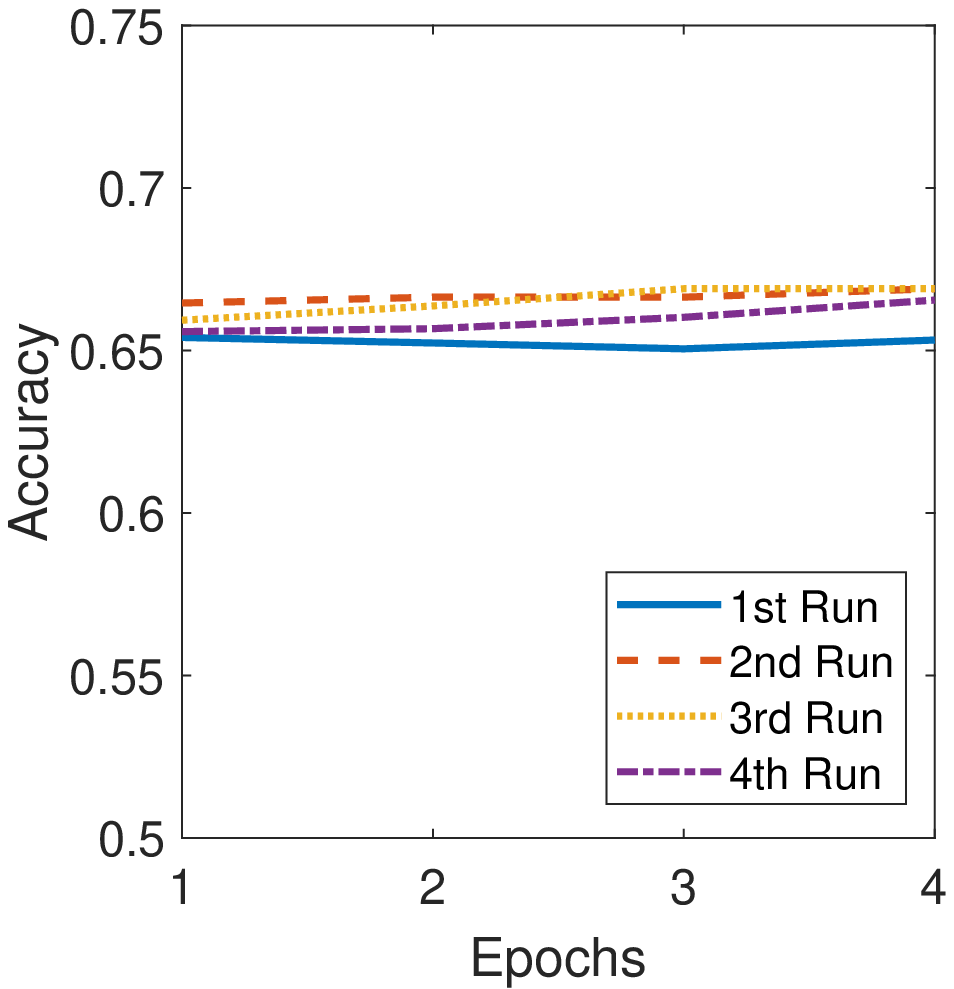}}
\subfloat[W=4]{\includegraphics[width=0.25\textwidth]{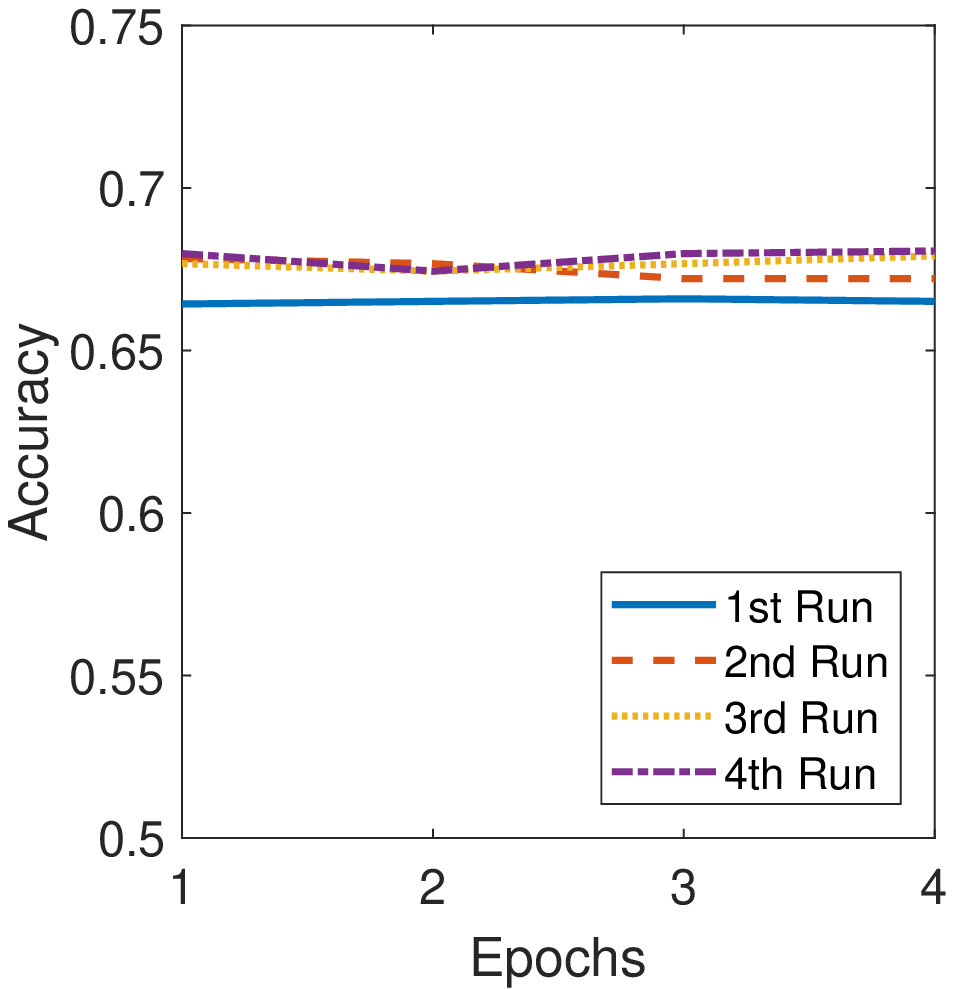}}
\caption{Classification performance with respect to the epochs of the algorithm for each run (Sparsity Level=3).}
\label{fig12}
\end{figure*}

Fig. \ref{fig13} demonstrates the distribution of two dictionaries (we used the dictionaries that were produced from the $4^{th}$ run of our algorithm) that refer to the spontaneous neuronal activity of the 9-day old and the 36-day old mouse under the consideration of $W=3$ and $SL=3$. The figure demonstrates the number of various patterns (doublets, triplets etc.) co-activating within a temporal window of 3 time bins. As in the case of $W=1$, we observe that the number of the various synchronicity patterns for the 9-day old mouse is greater than the number of the patterns for the 36-day old mouse. Additionally, the dictionary that refers to the activity of the 9-day old mouse includes more complex patterns with more than 20 neurons having a temporal correlation within 3 time bins, while such patterns appear in much smaller numbers for the 36-day old mouse. Finally, the size of each dictionary also reveals information about the data that we summarize. The dictionary that refers to the activity of the 9-day old mouse has greater size than the dictionary that refers to the activity of the 36-day old mouse, which correctly indicates and verifies that it fires less.

\subsection{Time Complexity and Convergence Analysis}
The main computational burden of ADL lies in the calculations of the coefficient matrices. ADL performs 4 epochs, and each epoch executes a number of iterations equal to the number of training examples (Section II, Fig. \ref{fig3}). To compute the reconstructed signals $\mathbf{V}_{clean\_reconstructed}$, $\mathbf{V'}_{clean\_reconstructed}$, $\mathbf{V}_{noisy\_reconstructed}$ and $\mathbf{V'}_{noisy\_reconstructed}$ at each iteration, so that they are used in the calculation of the representation errors $\mathbf{E}_{clean}$, $\mathbf{E'}_{clean}$, $\mathbf{E}_{noisy}$ and $\mathbf{E'}_{noisy}$ respectively, our algorithm calculates 4 coefficient matrices using the OMP method. The time complexity of OMP at a given iteration $T_0$ is $O(kM^2+k+T_0M+{T_0}^2)$ \cite{mailhe2009low}, where $k$ is the number of dictionary atoms, $M$ is the dimension of the signal and $T_0$ indicates the number of atoms that have been selected (i.e. the sparsity level). Thus, given the fact that in every iteration of our algorithm, OMP is calculated 4 times and our algorithm executes 4 epochs, the cost of our method is $O(16N_1(kM^2+k+T_0M+{T_0}^2))$, where $N_1$ is the number of iterations of ADL. Regarding the time complexity of the RADL algorithm, after we obtain the dictionary $D$ and before the beginning of a new epoch, RADL uses the set ${\bf T}_1$ to keep only those dictionary elements, which are mostly used for the representation of this set (Section II). Thus, the time complexity of the RADL is $O(16N_2(kM^2+k+T_0M+{T_0}^2)+4k)$, where $N_2$ is the number of iterations of RADL. Notice that in the case of the RADL algorithm, it has a larger set of training examples ($N_2>N_1$), which results to more iterations, and thus to a higher time complexity.

\begin{figure}[!h]
\centering
\subfloat[W=1]{\includegraphics[width=0.25\textwidth]{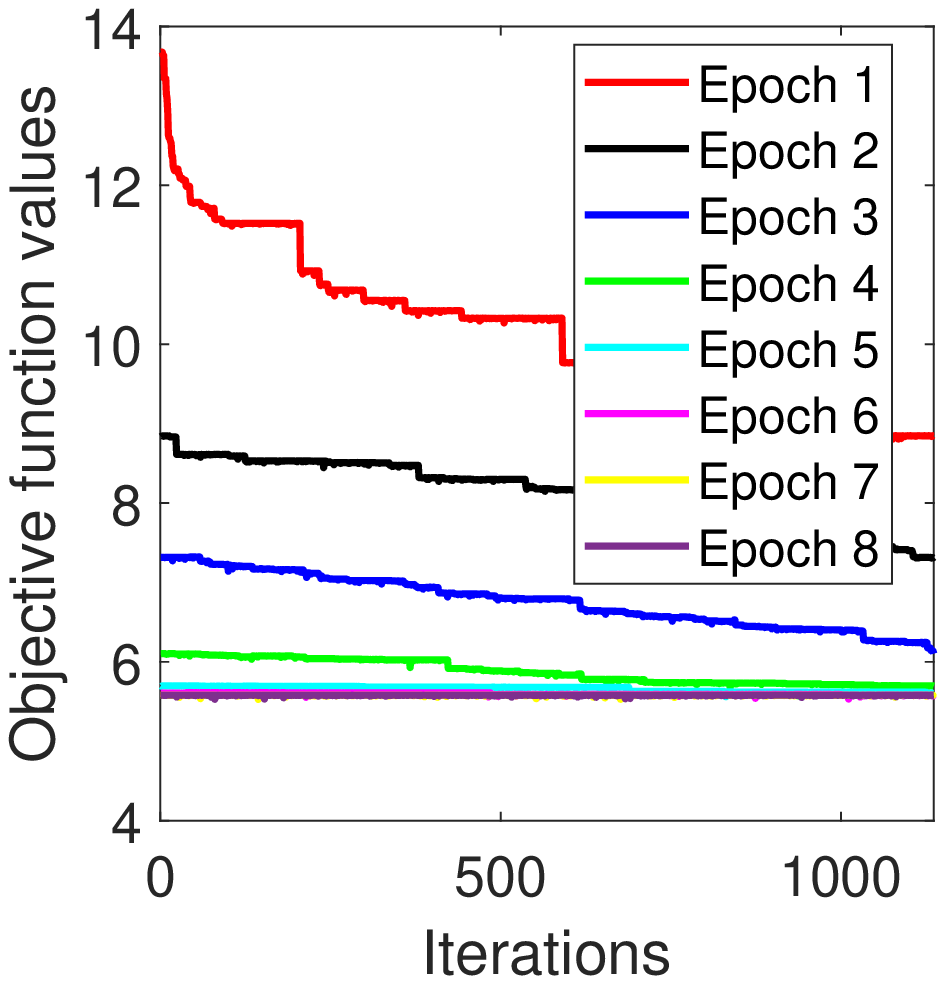}}
\subfloat[W=2]{\includegraphics[width=0.25\textwidth]{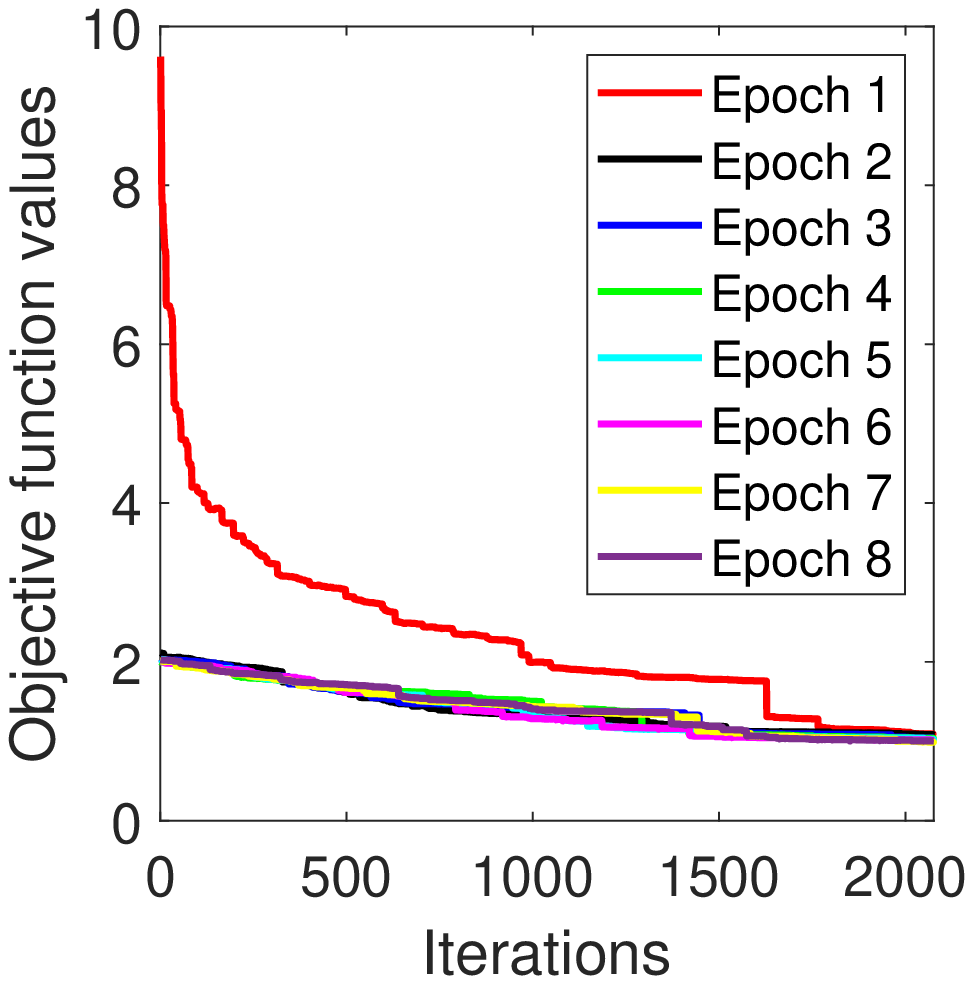}}
\caption{Convergence curves of the objective function values versus iterations for (a) ADL (W=1) and (b) RADL (W=2)}
\label{fig15}
\end{figure}

Concerning the convergence nature of the algorithms, we report in Fig. \ref{fig15} the objective function values of ADL as well as of RADL (W=2) with respect to the number of iterations of the algorithms, when they execute $8$ epochs. We observe that the objective function values of both algorithms are non-increasing during the iterations, and they both converge to a small value. Compared to ADL, RADL converges faster and to a lower value than ADL. It also needs only one epoch for that, while ADL reaches its lowest value in the $4^{th}$ epoch.  

\section{Related Work}
The past several years have witnessed the rapid development of the theory and algorithms of DL methods. DL has been successfully applied in various domains, such as image classification and denoising, remote sensing, face recognition, digit and texture classification etc. The success of these methods lie in the fact that high-dimensional data can be represented or coded by a few representative samples in a low-dimensional manifold.

In remote sensing, Li \textit{et al.} \cite{li2014recovering} addressed the problem of cloud cover and accompanying shadows, which are two of the most common noise sources for the majority of remote sensing data in the range of the visible and infrared spectra. For the recovery of surficial information, which is very important for target recognition, classification, segmentation etc, they proposed two multitemporal DL algorithms, expanding on their K-SVD and Bayesian counterparts. Li \textit{et al.} \cite{li2015sparse} also addressed the problem that remote sensing images are easily subjected to information loss, due to sensor failure and poor observation conditions. Thus, they proposed an analysis model for reconstructing the missing information in remote sensing images, so that more effective analysis of the earth can be accomplished. 

In image and video processing, where it is common to learn dictionaries adapted to small patches, with training data that may include several millions of these patches, Mairal \textit{et al.} \cite{mairal2009online} proposed an online dictionary learning (ODL) algorithm based on stochastic approximations, which scales up to large datasets with millions of training samples handling also dynamic training data changing over time, such as video sequences. In the same context of image processing, Iqbal \textit{et al.}  \cite{iqbal2019alpha} proposed a DL algorithm, which minimizes the assumption on the noise by using a function derived from the $\alpha$-divergence, which is used in the data fidelity term of the objective function instead of the quadratic loss or the Frobenius norm. The algorithm is applied on various image processing applications, such as digit recognition, background removal, and grayscale image denoising. 

For the task of face as well as object recognition, Li \textit{et al.} \cite{li2019discriminative} proposed a discriminative Fisher embedding DL algorithm to concurrently preserve both interclass variances and intraclass similarities of the learned dictionary and coding coefficients in the embedding space. One of the first successful attempts in discriminative DL was the Discriminative K-SVD (D-KSVD) algorithm \cite{zhang2010discriminative} for face recognition. They extended K-SVD by incorporating the classification error into the objective function, thus allowing the performance of a linear classifier and the representational power of the dictionary to be considered at the same time in the same optimization procedure, while in our work these are considered two seperate steps (i.e. classification error is not incorporated in the objective function). In several variants of discriminative DL methods are proposed to improve the data representation and classification abilities by encoding the locality and reconstruction error into the DL procedures, while some of them aim to concurrently improve the scalability of the algorithms by getting rid of costly norms \cite{zhang2019joint,zhang2017jointly,zhang2019scalable}. Recently, DL has also been extended to deep learning frameworks \cite{mahdizadehaghdam2019deep}, which seek multiple dictionaries at different image scales capturing also complementary coherent characteristics. 

\section{Conclusions and Future Work}
In this work we proposed the Adversarial Dictionary Learning algorithm (ADL) that was applied on real-world data that refer to the spontaneous neuronal activity of a 9-day old and a 36-day old mouse over time. In order to examine the extent to which the trained dictionary had captured the underlying statistics of the data, we trained and tested an SVM-classifier with the reconstructed clean and noisy signals that were produced from our method as well as with the reconstructed signals produced from other dictionary learning methods. The results on the classification accuracy showed that our method can better discriminate the true from the noisy activation patterns, indicating the robustness of our method. Moreover, in contrast to other dictionary learning methods, our framework also produces an interpretable dictionary, consisting only with the most robust activation patterns of the input data and not with real-numbered values, which have no physical meaning. We also extended the idea of ADL to a more relaxed approach, proposing thus the RADL algorithm, which produces a dictionary that captures patterns within bigger time window intervals and is not restricted to the synchronous activity of neurons within the same time bin. Experimental results demonstrate that increasing the activation patterns time window, has a positive effect on the classification performance.

Future work will focus on the extension of our algorithm with graph signal processing methods, which could provide insights related to the temporal dynamics of the network as well as its functional network activities. We also plan to explore the potential of the proposed method in characterizing normal brain organizations as well as alterations due to various brain-disorders, such as schizophrenia, autism, and Alzheimer's disease.

\section*{Acknowledgment}
\noindent This research is co-financed by Greece and the European Union (European Social Fund-ESF) through the Operational Programme “Human Resources Development, Education and Lifelong Learning” in the context of the project “Strengthening Human Resources Research Potential via Doctorate Research” (MIS-5000432), implemented by the State Scholarships
Foundation (IKY). S.M.S. received support from the Simons Foundation SFARI Research Award No. 402047, NEI Grant RO1-EY-109272, and NINDS R21 NS096640. This work is also supported from the Hellenic Foundation for Research and Innovation (HFRI) and the General Secretariat for Research and Technology under grant agreement No. 2285, Erasmus+ International Mobility between University of Crete and Harvard Medical School 2017-1-EL01-KA107-035639, and the Marie Curie RISE NHQWAVE project under grant agreement No. 4500.



\bibliography{refs}

\end{document}